\documentclass[aps,twocolumn] {revtex4-1}
\usepackage{amsmath,graphicx,amsfonts}

\usepackage{times}
\usepackage[colorlinks=true,linkcolor=blue,urlcolor=blue,citecolor=blue,breaklinks=true]{hyperref}

\usepackage[utf8]{inputenc}
\DeclareUnicodeCharacter{03BC}{\mbox{$\mu$}}
\DeclareUnicodeCharacter{2009}{ }

\def\vep{\varepsilon}

\def\wt{\widetilde}

\def\D{{\mathcal{D}}}

\newcommand{\be}{\begin{equation}}
\newcommand{\ee}{\end{equation}}

\def\eq#1{Eq.~(\ref{#1})}
\def\la{\langle}
\def\ra{\rangle}

\def\vk#1{{{\bf{k}}_{#1}}}
\def\Gi#1{{\Gamma_1({{k}}_{#1})}}
\def\Gt#1#2{{\Gamma_3^\theta(k_{#1},k_{#2})}}
\def\Gs#1#2#3{{\Gamma_3^\sigma(k_{#1},k_{#2},k_{#3})}}
\def\Gf#1#2#3#4{{\Gamma_4(k_{#1},k_{#2},k_{#3},k_{#4})}}

\def\da{\dagger}

\def\wt{\widetilde}
\def\om{\omega}

\def\ao#1{\la{#1}\ra}
\def\ac#1{\la{#1}\ra_c}
\def\si#1{\sigma_{{\bf k}_{#1}}}
\def\te#1{\theta_{{\bf k}_{#1}}}

\def\da{\dagger}
\def\D{{\mathcal{D}}}

\date\today
\begin{document}
	
\title{Field-theoretical approach to the Casimir-like interaction in a one-dimensional Bose gas}
\author{Benjamin Reichert}
\author{Aleksandra Petkovi\'{c}}
\author{Zoran Ristivojevic}
\affiliation{Laboratoire de Physique Th\'{e}orique, Universit\'{e} de Toulouse, CNRS, UPS, 31062 Toulouse, France}
	
\begin{abstract}
We study the fluctuation-induced interaction between two impurities in a weakly-interacting one-dimensional Bose gas using
the field theoretical approach. At separations between impurities shorter and of the order of the healing length of the system, the induced interaction has a classical origin and behaves exponentially. At separations longer than the healing length, the interaction is of a quantum origin and scales as the third power of the inverse distance. Finite temperature destroys the quasi-long-range order of the Bose gas and, accordingly, the induced interaction becomes exponentially suppressed beyond the thermal length. We obtain analytical expressions for the induced interaction at zero and finite temperature that are  valid at arbitrary distances. We discuss experimental realizations as well as possible formation of bound states of two impurities, known as bipolarons.
\end{abstract}

\maketitle
%%%%%%%%%%%%%
\section{Introduction}
%%%%%%%%%%%%%

One of the most peculiar predictions of quantum field theory is that  the vacuum  should  be considered as a fluctuating entity. In 1948,  Casimir predicted that two neutral large conducting plates placed in a vacuum should  attract each other with a force depending on their relative separation \cite{casimir1948}. This electromagnetic Casimir effect is a result of the modification of the ground state energy  due to constrains on quantum fluctuations imposed by the plates. Early experimental attempts to show the existence of such effective interaction were only in qualitative agreement with the theoretical predictions \cite{bordag_advances_2009}. It is rather recently that accurate quantitative results confirmed the predictions of Casimir \cite{lamoreaux_demonstration_1997,mohideen_precision_1998,klimchitskaya_casimir_2009}. More generally, the Casimir-like effect is studied in different fields of physics and refers to  an effective interaction between external objects placed in a fluctuating medium. The effective interaction depends on the correlations of fluctuations. In superfluids for instance, where correlations are long-ranged, the effective interaction is also expected to be long-ranged \cite{kardar_friction_1999}. 

Thanks  to  experimental progresses made in the manipulation of ultracold atoms, where various superfluids can now be realized \cite{bloch_many-body_2008}, the interest in  the effective Casimir-like  interaction between impurities in a quantum liquid has been renewed \cite{bardeen_effective_1967,bijlsma_phonon_2000,roberts_casimir-like_2005,klein_interaction_2005,recati_casimir_2005, fuchs_oscillating_2007,wachter_indirect_2007,schecter_phonon-mediated_2014,marino_casimir_2017,dehkharghani_coalescence_2018,yu_casimir_2009,reichert_casimir_2018,pavlov_phonon-mediated_2018}. 
The case of a  one-dimensional weakly interacting Bose gas has been studied in Refs.~\cite{klein_interaction_2005,recati_casimir_2005,schecter_phonon-mediated_2014,dehkharghani_coalescence_2018,reichert_casimir_2018}. Even though the system exhibits quasi long-range correlations due to its superfluid nature, in Refs.~\cite{klein_interaction_2005,recati_casimir_2005,dehkharghani_coalescence_2018} was found a short-ranged attraction that decays exponentially with a distance. Recently, in  Ref.~\cite{schecter_phonon-mediated_2014} it was shown that the effective interaction is of a long-range nature. Namely, it was predicted that it decays algebraically as the third power of the inverse distance between impurities \cite{schecter_phonon-mediated_2014}.

In Ref.~\cite{reichert_casimir_2018} we were able to resolve this discrepancy present in the literature by providing an analytic expression for the induced interaction at zero-temperature  valid at arbitrary distances between the impurities. Our study was based on a microscopic approach which accounted for the quantum corrections to the mean-field Gross-Pitaevskii equation, which naturally takes into consideration the nonlinearity of the excitation spectrum. 
We found that at short distances the interaction decays exponentially as predicted in    Refs.~\cite{klein_interaction_2005,recati_casimir_2005,dehkharghani_coalescence_2018}, while at large distances it decays algebraically as found in Ref.~\cite{schecter_phonon-mediated_2014}. We calculated analytically the  interaction in the crossover region between these two limiting cases. 

In this paper we revisit the problem of the induced interaction between the impurities in a one-dimensional Bose gas using a complementary method, which is a systematic field theoretical approach based on the path integral in imaginary time. We use a hydrodynamic description of the system where the density and the phase become the relevant degrees of freedom. The central quantity of our interest is the behavior of the Helmholtz free energy of the Bose gas as a function of the distance between the impurities. It is calculated here at one-loop order using the perturbation theory at weak interaction. We considered both cases of zero and finite temperature, in contrast to Ref.~\cite{reichert_casimir_2018}. At zero temperature we find a result that is identical to the one found in Ref.~\cite{reichert_casimir_2018}. At finite temperatures, we obtain the result for the induced interaction valid at arbitrary distances and in a wide range of temperatures. Thermal fluctuations tend to destroy the quasi long-range order in a one-dimensional superfluid and accordingly modify the quantum fluctuation-induced long range interaction turning it into an exponential one, which is in agreement with the general picture \cite{kardar_friction_1999}.

The paper is organized as follows. In Sec.~\ref{sec:Model} we introduce the model and the field-theoretical method used to describe the system. In Sec.~\ref{sec:Free} we then develop the perturbative framework, which we use in Sec.~\ref{sec:4} to calculate the Landau free energy of the system to one-loop order. In Sec.~\ref{sec5}, by performing a Legendre transformation of the Landau free  energy we obtain the Helmholtz free energy. We then study its behavior with respect to the distance between impurities, and obtain the induced interaction in Sec.~\ref{sec:Eff}. We study both cases of zero and finite temperatures. Section \ref{sec:Conclusion} contains the discussions and conclusions. Some more technical details and side results are presented in Appendices.

%%%%%%%%%%%%%
\section{Model}
%% %%%%%%%%%%%
\label{sec:Model}

We study one-dimensional bosons with contact repulsion in the presence of two impurities. The system is described by the Hamiltonian
\begin{align}\label{eq:h0}
H={}&\int {d}x\left[\frac{\hbar^2}{2m}(\nabla \hat\psi^\da)(\nabla \hat\psi)+\frac{g}{2} \hat\psi^\da\hat\psi^\da\hat\psi\hat\psi\right] \notag\\
&+G\left[\hat n\left(-\ell/2\right)+\hat n\left(\ell/2\right)\right], \end{align}
where $\hat\psi^\da$ and $\hat\psi$ are the bosonic single-particle operators that satisfy the standard commutation relation $[\hat\psi(x),\hat\psi^\da(x')]=\delta(x-x')$. In Eq.~(\ref{eq:h0}), $m$ is the mass of bosonic particles, $g>0$ denotes the strength of repulsive contact interaction between the bosons while $\hbar$ is the reduced Planck constant. The two static impurities at positions $\pm \ell/2$ locally couple to the density of the Bose liquid $\hat n=\hat{\psi}^\da\hat{\psi}$, where $G$ is the coupling strength. At $G=0$, the Hamiltonian (\ref{eq:h0}) corresponds to the integrable Lieb-Liniger model \cite{lieb_exact_1963}.

Our goal is to compute the interaction energy between the two impurities mediated by bosonic excitations.  In order to perform this calculation we will evaluate the Helmholtz free energy. We treat analytically the weakly-interacting Bose liquid using a hydrodynamic description. We represent the single-particle operators as
\begin{align}
\hat\psi^\da=\sqrt{\hat n}e^{i\hat\theta},\quad \hat\psi=e^{-i\hat\theta}\sqrt{\hat n},
\end{align}
where the density $\hat n$ and the phase $\hat\theta$ satisfy the bosonic commutation relation $[\hat n(x),\hat \theta(x')]=-i  \delta(x-x')$. The Hamiltonian (\ref{eq:h0}) can now be expressed as 
\begin{align}\label{H}
H={}&\frac{\hbar^2}{2m}\int d x\left[\hat n(\nabla\hat\theta)^2+\frac{(\nabla\hat n)^2}{4\hat n}\right]+\frac{g}{2}\int d x\, \hat n^2\notag\\
&+G\left[\hat n(\ell/2) +\hat n(-\ell/2)\right]+h_0,
\end{align}
where
\begin{subequations}\label{4}
\begin{align}\label{4a}
h_0=-\Delta\mu\int d x  \,\hat n+L\Delta \epsilon,
\end{align}	
\begin{align}\label{4b}
\Delta \mu=\frac{g}{2}\delta(x)|_{x\to 0},\quad \Delta \epsilon =\frac{\hbar^2}{4m}\nabla^2\delta(x)|_{x\to 0}.
\end{align}
\end{subequations}
In the Hamiltonian (\ref{H}) we accounted for the two terms given by Eq.~(\ref{4a}) that are formally divergent. They are usually neglected in studies of the excitations of the system. However, they are important for the ground state energy. The terms in $h_0$ arise from the commutation relations once one transforms the Hamiltonian (\ref{eq:h0}) into the form (\ref{H}). Notice that $h_0$ is formally divergent due to presence of  delta functions and need to  be regularized, e.g., by the substitution
$\delta(x)\to (1/L)\sum_{|k|<\Lambda} e^{i k x}$, 
where $L$ is the size of the system, while $\Lambda$ is a high-momentum cutoff.

We now employ the path integral formalism \cite{braaten_quantum_1999,andersen_theory_2004} in order to evaluate the Landau free energy of the system $F=-\ln Z/\beta$. Here $\beta=1/T$, where $T$ is the temperature, while $Z$ is the partition function. We set the Boltzmann constant to unity. The partition function is given as the functional integral
\begin{align}\label{Z}
Z=\int \D n(x,\tau)\D\theta(x,\tau)e^{-S[n(x,\tau),\theta(x,\tau)]},
\end{align}
where the (dimensionless) action is 
\begin{align}\label{S}
S=\frac{1}{\hbar}\int_0^{\hbar \beta}d\tau\int d x (\psi^*\hbar\partial_\tau\psi+\mathcal{H}-\mu n).
\end{align} 
By $\mu$ we denote the chemical potential of the system, $\mathcal{H}$ is the Hamiltonian density, while the complex field
\begin{align}
\label{eq:nt}
\psi(x,\tau)=\sqrt{n(x,\tau)}e^{-i\theta(x,\tau)}
\end{align} 
is periodic in imaginary time $\psi(x,0)=\psi(x,\hbar\beta)$.

Since we consider the weakly-interacting Bose gas and weakly-coupled impurities, we can assume that the density fluctuations are small. We thus parametrize the density as
\begin{align}\label{n}
n(x,\tau)=\sigma_0+\sigma(x,\tau),
\end{align}
where the field $\sigma(x,\tau)$ accounts for the density fluctuations around some constant value $\sigma_0$. We emphasize that we are working within the formalism of the grand canonical ensemble and that $\sigma_0\equiv \sigma_0(\mu)$ is a function of the chemical potential, which will be be determined in a variational manner, as detailed in Sec.~\ref{sec:Free}. Since $|\sigma|\ll \sigma_0$, we can expand the square root in Eq.~\eqref{eq:nt} and then regroup the terms of different smallness in the action (\ref{S}):
\begin{align}\label{Ssp}
S=S_c+S_0+S_1+S_3+S_4+\cdots
\end{align}
Here $S_c$ does not contain fluctuating fields; $S_0$ contains the quadratic terms in the fluctuating fields $\sigma$ and $\theta$, while $S_3$ and $S_4$ are the cubic and quartic anharmonic corrections, respectively. $S_1$ is linear in $\sigma$ and describes the interaction of the Bose gas with the impurities. We notice that under the parametrization (\ref{n}), the partition function (\ref{Z}) becomes dependent on $\sigma_0$:
\begin{align}
Z[\sigma_0]=\int \D \sigma(x,\tau)\D\theta(x,\tau) e^{-S[\sigma_0,\sigma(x,\tau),\theta(x,\tau)]}.
\label{Zs}
\end{align}

The part of the action $S_c$ could be conveniently split as $S_c=S_c^{(0)}+S_c^{(1)}$, where $S_c^{(0)}$ is the classical action. They take the form  
\begin{gather}\label{Sc}
S_c^{(0)}=\beta L (g\sigma_0^2/2-\mu\sigma_0)+2\beta G\sigma_0,\\
\label{Sc1}
S_c^{(1)}=\beta L [-\Delta\mu\sigma_0+\Delta\epsilon].
\end{gather}
By $L$ we denote the system size. We notice that the term $h_0$ from the Hamiltonian (\ref{H}) produces the correction $S_c^{(1)}$.

The quadratic part of the action describes the dynamics of free excitations and is given by
\begin{align}\label{eq:S0xt}
S_0=\int  d\tau   dx\left\{\dfrac{\hbar\sigma_0}{2m}\left[(\nabla\theta)^2+\dfrac{(\nabla\sigma)^2}{4\sigma_0^2} \right]+\dfrac{g \sigma^2}{2\hbar}-i\sigma\partial_\tau\theta\right\}. 
\end{align}
Thanks to the invariance under the translation and the periodicity in the imaginary time of $S_0$, it is convenient to work in Fourier space where the fluctuating fields $\sigma$ and $\theta$ are given by 
\begin{subequations}\label{eq:fts}
\begin{align}
\sigma(x,\tau)=\dfrac{1}{\sqrt{L\hbar\beta }}\sum_{\vk{}}e^{i(kx-\om \tau)}\si{},
\end{align}
\begin{align}
\theta(x,\tau)=\frac{1}{\sqrt{L\hbar\beta }}\sum_{\mathbf{k}}e^{i(kx-\omega \tau)}\te{},
\end{align}
\end{subequations}
with ${\bf{k}}=(k,\omega)$ and where $\om\equiv\om_j =2\pi j/\hbar \beta$ ($j\in \mathbb Z$) is the bosonic Matsubara frequency.
Since the density and phase fields are real in real space, one has $\sigma_{-\vk{}}=\si{}^*$ and $\theta_{-\vk{}}=\te{}^*$ in Fourier space. We point out that in these summations, the  $k=0$ mode  is removed such that all the $x$-independent part is contained in $\sigma_0$ [cf.~Eq.~\eqref{n}]. In principle one should impose a cutoff $\Lambda$ on the momentum summation in order to prevent large fluctuations. It will be, however, explicitly  displayed in the sums only when needed to regularize the divergent momentum summations. In Sec.~\ref{sec:Free}, the divergent terms are absorbed in the renormalization procedure and the limit $\Lambda\to\infty$ then  can be safely taken.

Using Fourier decomposition (\ref{eq:fts}), the quadratic action (\ref{eq:S0xt}) becomes
\begin{align}\label{S0}
S_0=\frac{1}{2}\sum_\mathbf{k,q} (\sigma_\mathbf{k},\theta_\mathbf{k}) G^{-1}_\mathbf{k,q} \left(\sigma_\mathbf{q}\atop\theta_\mathbf{q}\right).
\end{align}
The propagator $G$ encodes the correlation functions
\begin{align}\label{G}
G_{\mathbf{k,q}}&=\left( \begin{matrix} 
\langle\sigma_\mathbf{k} \sigma_\mathbf{q}\rangle& \langle\sigma_\mathbf{k} \theta_\mathbf{q}\rangle\\
\langle\theta_\mathbf{k} \sigma_\mathbf{q}\rangle & \langle\theta_\mathbf{k} \theta_\mathbf{q}\rangle
\end{matrix} \right)\notag\\ &=\frac{\hbar}{\hbar^2\omega^2+\varepsilon_k^2} \left( \begin{matrix} 
\frac{\hbar^2k^2\sigma_0}{m}& -\hbar\omega \\
\hbar\omega  & \frac{m\varepsilon_k^2}{\sigma_0\hbar^2k^2}
\end{matrix} \right)\delta_{\mathbf{k+q},0}.
\end{align}
Here the average $\langle\ldots\rangle$ is with respect to the quadratic action
\begin{align}\label{Amean}
\langle A\rangle =\frac{1}{Z_0}\int\mathcal{D}\theta\mathcal{D}\sigma A e^{-S_0[\sigma_0,\sigma,\theta]},
\end{align}
where 
\begin{align}\label{Z0}
Z_{0}=\int\mathcal{D}\theta\mathcal{D}\sigma e^{-S_0[\sigma_0,\sigma,\theta]}.
\end{align}
The spectrum of elementary excitations in Eq.~(\ref{G}) has the Bogoliubov form
\begin{align}\label{eq:Bogspectrum}
\varepsilon_k=\sqrt{g\sigma_0\frac{\hbar^2k^2}{m}+\left(\frac{\hbar^2k^2}{2m}\right)^2}.
\end{align} 
Notice that the nonlinearity appearing in the Bogoliubov spectrum is due to the quantum pressure term $\propto (\nabla\sigma)^2$ in Eq.~(\ref{eq:S0xt}). This term is usually discarded in a low-energy description of the system, e.g., in a Luttinger liquid. However this term is of crucial importance for the correct description of the effective interaction between impurities at arbitrary distances.

In a consistent theory one must account for the anharmonic terms that describe the interaction between Bogoliubov quasiparticles \cite{gangardt_bloch_2009,ristivojevic_decay_2016}. The cubic and quartic anharmonic terms in the action $S$ are, respectively, 
\begin{gather}
S_3={}\dfrac{\hbar}{2m}\int  d\tau   dx\left[ \sigma(\nabla \theta)^2-\sigma\dfrac{(\nabla\sigma)^2}{4\sigma_0^2}\right],\\
S_4={}\dfrac{\hbar}{8m\sigma_0^3}\int  d\tau   dx{\sigma^2(\nabla\sigma)^2}\label{eq:S4xt}.
\end{gather}
In Fourier space, $S_3$ becomes
\begin{gather}
S_3=\sum_{\mathbf{k},\mathbf{q},\mathbf{p}} \bigl[\Gamma_{3}^\theta(q,p)\sigma_{\mathbf{k}} \theta_{\mathbf{q}} \theta_{\mathbf{p}}
+\Gamma_{3}^\sigma(k,q,p)\sigma_{\mathbf{k}} \sigma_{\mathbf{q}} \sigma_{\mathbf{p}}\bigr]\delta_{\mathbf{k}+\mathbf{q}+\mathbf{p},0},\notag\\
\Gamma_{3}^\theta(q,p)=-\frac{\sqrt{\hbar}}{\sqrt{\beta L}}\frac{ qp}{2m},\notag\\ \Gamma_{3}^\sigma(k,q,p)=-\frac{\sqrt{\hbar}}{\sqrt{\beta L}}\frac{k^2+q^2+p^2}{48m\sigma_0^2},
\end{gather}
while $S_4$ takes the form
\begin{gather}
S_4=\sum_{\mathbf{k},\mathbf{q},\mathbf{p},\mathbf{r}} 
\Gamma_{4}(k,q,p,r)\sigma_{\mathbf{k}} \sigma_{\mathbf{q}} \sigma_{\mathbf{p}}\sigma_{\mathbf{r}}\delta_{\mathbf{k}+\mathbf{q}+\mathbf{p}+\mathbf{r},0},\notag\\
\Gamma_{4}(k,q,p,r)=\frac{1}{\beta L}\frac{k^2+q^2+p^2+r^2}{96m\sigma_0^3}.
\end{gather}
The part of the action that describes the interaction with the impurities has the form
\begin{align}
S_1={}&\frac{G}{\hbar}\int d\tau \left[\sigma(\ell/2,\tau)+\sigma(-\ell/2,\tau)\right]. \label{eq:S1xt}
\end{align}
Since $\sigma_{\bf{k}}=0$ at $k=0$, the term involving $\int dx \sigma(x,\tau) =0$ in $S$ vanishes. In Fourier space on thus has
\begin{align}\label{S1}
S_1=\sum_{\mathbf{k}} \Gamma_1(k)\sigma_\mathbf{k}\delta_{\omega,0},\quad \Gamma_1=\frac{2\sqrt{\beta}}{\sqrt{\hbar L}} G\cos(k\ell/2).
\end{align}
At weak interaction, one can use the scaling analysis \cite{ristivojevic_decay_2016} to show that $S_0$, $S_3$, and $S_4$ correspond to the first three terms of the expansion of the action with respect to the small parameter ${\gamma}^{1/4}$, where $S_0\propto {\gamma}^0$, $S_3\propto {\gamma}^{1/4}$, and $S_4\propto \sqrt{\gamma}$.  Here $\gamma=mg/\hbar^2\bar n\ll 1$ and $\bar n$ is the mean boson density. At small boson-impurity coupling, $G\ll g/\sqrt{\gamma} $, one can show that the depletion of the boson density due to impurities is small. All this enables us to evaluate the partition function using the perturbation theory. Effectively, we perform two expansions, one in $\sqrt{\gamma}$ and another in $G\sqrt{\gamma}/g$ as we explain more precisely in the following sections. However the result for the induced interaction will cover all distances since we are dealing with asymptotically exact Bogoliubov dispersion (\ref{eq:Bogspectrum}).

%%%%%%%%%%%%%%%%%%%%%%%%
\section{Landau Free energy}
%%%%%%%%%%%%%%%%%%%%%%%%
\label{sec:Free}

In this section we explain how to formally evaluate the Landau free energy of the system using a perturbation theory at weak interaction. In the following sections we perform this calculation by  first finding the Landau free energy at the mean-field level and then we account for its leading quantum correction. We finally perform the Legendre transform to obtain the (Helmholtz) free energy that depends on the density.

The partition function of a system allows us to obtains all the thermodynamic quantities. In particular, the Landau free energy $F(\mu)$ is given by 
\begin{align}
\beta F(\mu)=-\ln Z,
\end{align}
where the partition function $Z$ is given by Eq.~(\ref{Z}). After introducing the parametrization (\ref{n}), the partition function becomes a functional of $\sigma_0$ [see Eq.~(\ref{Zs})]. The grand potential $\Omega(\mu,\sigma_0)$ is defined by the relation
\begin{align}
\beta \Omega(\mu,\sigma_0)=-\ln Z[\sigma_0],
\label{eq:Fdef}
\end{align}
where the partition function $Z[\sigma_0]$ is given by Eq.~(\ref{Zs}). Unlike the Landau free energy that depends on the chemical potential, the latter potential also depend on $\sigma_0$ [cf.~Eq.~(\ref{n})]. In the ground state, the physical value of  $\sigma_0$ should be the one which minimizes the grand potential \cite{andersen_theory_2004}. It is thus determined by the condition
\be
\left.\dfrac{\partial \Omega(\mu,\sigma_0)}{\partial \sigma_0}\right|_{\sigma_0=\bar n}=0,\label{eq:sigmabar}
\ee
where $\bar n\equiv\bar n(\mu)$ is actually a function of the chemical potential. Then the free energy is given by the grand potential evaluated at $\sigma_0=\bar n$:
\be
F(\mu)=\Omega(\mu,\bar n).
\label{eq:FOm}
\ee
The quantity $\bar n$ corresponds to the mean density of the system,
and is defined by the relation 
\begin{align}\label{eq:density}
\bar n=-\frac{1}{L}\frac{\partial F}{\partial\mu}.
\end{align}
Therefore the calculation of the grand potential $\Omega(\mu,\sigma_0)$ allows us to obtain the Landau free energy. However obtaining an exact result involving the full partition function $Z[\sigma_0]$ is a very difficult task. Nevertheless, since we consider a weakly-interacting system  a perturbative expansion in the interaction of the grand potential can be performed.

Let us now  develop the perturbative scheme used in the reminder of the paper. We  start by using Eqs.~(\ref{Ssp}) and (\ref{Zs}) in \eq{eq:Fdef} to express the  grand potential as  
\begin{align}
\beta\Omega(\mu,\sigma_0)=S_c-\ln Z_0 -\ln\la e^{-(S_1+S_3+S_4)} \ra,
\label{eq:Z_Sint}
\end{align}
where we remind the reader that the definition of $\ao{\ldots}$ is given by Eq.~(\ref{Amean}). The challenging part here is the calculation of $\ln\la e^{-(S_1+S_3+S_4)} \ra$. It can be expressed using the cumulant expansion as
\begin{align}
\ln\la e^{-O} \ra=\sum_{n=1}^\infty\frac{(-1)^n}{n!}\ac{O^n}, \label{eq:connected}
\end{align}
where $\ac{\ldots}$ denotes the  connected cumulants evaluated with respect to $S_0$.  The first several connected cumulants are given by \cite{ma}
\begin{align}
\ac{A}={}&\ao{A},\\
\ac{A^2}={}&\ao{A^2}-\ao{A}^2,\\ \ac{A^3}={}&\ao{A^3}+2\ao{A}^3-3\ao{A}\ao{A^2},\\ \ac{A^4}={}&\ao{A^4}-6\ao{A}^4-3\ao{A^2}^2-4\ao{A}\ao{A^3}\notag\\
&+12\ao{A}^2 \ao{A^2}.
\end{align}
 The different connected cumulants can be classified as a function of the number of loops contained in them. A loop is defined as an unconstrained sum over a momentum which is not directly involved in $\Gamma_1$ [see Eq.~(\ref{S1})]. For example, the integral over $q$ in $\int dk dq \Gamma_1^2(k)f(k,q)$, where $f$ is an arbitrary function of $k$ and $q$, would be considered as a loop integral while the integral over $k$ would not. 
One can show that each loop brings a factor $\sqrt{\gamma}$. Therefore it is convenient  to express the grand potential as the  loop expansion
\be
\Omega(\mu,\sigma_0)= \Omega_0(\mu,\sigma_0)+ \Omega_1(\mu,\sigma_0)+\ldots,
\ee
where $\Omega_n$ is the $n$-loops contribution to the grand potential.
Due to  the smallness of Lieb's parameter $\gamma$ in weakly interacting Bose gas, higher loop contributions bring smaller and smaller quantum corrections to the zero-loop contribution that is also known as the mean-field contribution. For the purpose of this paper it is sufficient to calculate $\Omega$ to one-loop order. 

The Landau free energy $F(\mu)$ [\eq{eq:FOm}] can also be expressed as an expansion in quantum corrections:
\be\label{eq:expansionF}
F(\mu)=F_0(\mu)+F_1(\mu)+\ldots.
\ee
However we emphasize that the loop expansion $\Omega_0(\mu,\bar n)+\Omega_1(\mu,\bar n)+\ldots$ and the expansion in quantum corrections (\ref{eq:expansionF}) do not coincide since the density  $\bar n$ can be itself also expanded into loops as 
\be
\bar n=\bar n_0+\bar n_1+\ldots.
\ee
Here the mean density $\bar n_0$ is the minimum of the zero-loop grand potential. It is is defined via the relation 
\be
\left.\dfrac{\partial\Omega_0(\mu,\sigma_0)}{\partial \sigma_0}\right|_{\sigma_0=\bar n_0}=0. \label{eq:barn0}
\ee
By $\bar n_1$ is denoted the first quantum correction to the density. One can show  that the first two contributions in the expansion of $F(\mu)$ in quantum corrections are given by
\begin{gather}
F_0(\mu)=\Omega_0(\mu,\bar n_0), \label{eq:F0O0}\\
F_1(\mu)=\Omega_1(\mu,\bar n_0). \label{eq:F1O1}
\end{gather}
We loosely call these two terms the zero- and one-loop free energy.

%%%%%%%%%%%%%%%%%%%%%%%%%%%
\section{Free energy in loop expansion}\label{sec:4}
%%%%%%%%%%%%%%%%%%%%%%%%

\subsection{Zero-loop free energy}

We start by the evaluation of the zero-loop free energy.  As seen previously,  one needs to evaluate the zero-loop grand potential first. It is given by
\be
\beta \Omega_0(\mu,\sigma_0)=S_c^{(0)}-\dfrac{1}{2}\ac{(S_1)^2}+\ldots\label{eq:w0}
\ee
where  $\ldots$ stands for zero-loop terms containing higher powers of $S_1$. For the purpose of this paper, it is sufficient to calculate the quadratic order term, $\propto G^2$, which is the lowest order at which one observes an effective interaction between impurities. The first term in the right-hand side of Eq.~(\ref{eq:w0}) is  simply  given by the classical action $S_c^{(0)}$ [Eq.~\eqref{Sc}]. This term accounts for the contribution of the Bose gas and the energy of the two uncorrelated impurities coupled with the Bose gas.
The second term of \eq{eq:w0} is given by
\begin{align}
-\frac{1}{2}\ac{(  S_1)^2}&=-\dfrac{1}{2} \sum_{\vk{},\vk{}'}\Gi{}\Gamma_1(k')\ao{\sigma_{\bf{k}} \sigma_{{\bf{k}}'}}\delta_{\om,0}\delta_{\om',0}\notag\\
&=-\beta{G^2}\sqrt{\dfrac{m \sigma_0}{\hbar^2g}}\left(1+e^{-\frac{2\ell \sqrt{m g \sigma_0}}{\hbar}}\right), \label{eq:S1S1}
\end{align}	
where we take the thermodynamic limit and replace $\sum_k(\cdots)\to(L/2\pi)\int dk(\cdots)$.
This contribution takes into account the correlation of the two impurities through real (as opposed to virtual)   density fluctuations. We thus obtain
\begin{align}\label{omega0}
\beta\Omega_0(\mu,\sigma_0)={}&\beta L \left(\frac{g\sigma_0^2}{2}-\mu\sigma_0\right)+2\beta G \sigma_0\notag\\
&-\beta{G^2}\sqrt{\dfrac{m \sigma_0}{\hbar^2g}} \left(1+e^{-\frac{2\ell \sqrt{m g \sigma_0}}{\hbar}}\right).
\end{align}
Now that $\Omega_0$ has been determined  to order $G^2$, one can  obtain  $\bar n_0$  from the minimization condition (\ref{eq:barn0}), which yields
\begin{align}
\bar n_0={}&\dfrac{\mu}{g}-\dfrac{2G}{gL}+\dfrac{G^2}{2gL  }\dfrac{1}{\xi_\mu\mu}\left[1+e^{-2\ell/\xi_\mu }\left(1 -\dfrac{2\ell}{\xi_\mu}\right) \right],
\label{eq:n0}
\end{align}
where we introduced the length scale 
\be
\xi_\mu=\hbar/\sqrt{m\mu}.\label{ximu}
\ee
Replacing $\sigma_0$ by $\bar n_0$ [\eq{eq:n0}] in the zero-loop grand potential (\ref{omega0}),  one obtains the zero-loop free energy  (\ref{eq:F0O0})   to order $G^2$:
\begin{align}
F_0(\mu)=-\dfrac{L\mu^2}{2g}+\dfrac{2G \mu}{g}-\dfrac{G^2}{g\xi_\mu}\left(1+ e^{-2\ell/\xi_\mu}\right).\label{eq:F0}
\end{align}
Here and for the reminder of this paper, we do not consider terms scaling as  $1/L$, $1/L^2$, etc. in the free energy since they vanish in the thermodynamic limit $L\to \infty$. Similarly, in the density, the terms scaling as $1/L^2$ and with higher powers of $1/L$ are discarded.

Calculating the density of the system using the Landau free energy (\ref{eq:F0}) shows that $\bar n_0$ is indeed the zero-loop density of the system:
\be
-\dfrac{1}{L}\dfrac{\partial F_0(\mu)}{\partial \mu}=\bar n_0.
\ee
Now, substituting $\sigma_0=\bar n_0$ into Eq.~(\ref{eq:Bogspectrum}) gives in the thermodynamic limit the well known Bogoliubov spectrum.

%%%%%%%%%%%%%%%%%%%%%
\subsection{One-loop free energy}
%%%%%%%%%%%%%%%%%%%%%

We now proceed with the evaluation  of the one-loop free energy. We first need to calculate the one-loop  grand potential. It is  given  to order $G^2$ by
\begin{align}
\beta \Omega_1(\mu,\sigma_0)={}&S_c^{(1)}-\ln Z_0+\ac{S_1S_3}\notag\\
&+\frac{1}{2}\ac{(S_1)^2S_4}-\frac{1}{4}\ac{(S_1)^2(S_3)^2}+\ldots,
\label{eq:Om1}
\end{align}
where  $\ldots$ stands for one-loop terms containing higher powers of $S_1$. Note that contributions $\ac{S_1S_4}=0$ and $\ac{S_1^2S_3}=0$, since they contain odd number of fluctuating fields $\sigma_k$.
We start by evaluating the contribution of  bosons to the grand potential, i.e., $ -\ln Z_0$, where $Z_0$ is given by Eq.~(\ref{Z0}). 
Since $S_0$ is quadratic, obtaining $Z_0$ only requires a Gaussian integration and  one obtains
\begin{align}
-\ln Z_0=\sum_{|k|<\Lambda}\dfrac{1}{2}\beta \vep_k +\sum_k\ln(1-e^{-\beta\vep_k}),\label{eq:Z0e}
\end{align}
where we performed the summation over the Matsubara frequencies and regularized the divergence at zero temperature by a cutoff $\Lambda$. It is instructive to isolate the divergent contributions in the limit $\Lambda\to\infty$ from this expression. Replacing the summation by the integral $\sum_k\to (L/2\pi)\int dk$ in the first term on the right hand side of Eq.~(\ref{eq:Z0e}), one has
\begin{align}
\dfrac{L}{4\pi}\int_{-\Lambda}^{\Lambda} dk \vep_k={}&\dfrac{L}{4\pi}\int_{-\Lambda}^{\Lambda} dk \left(\vep_k-{g \sigma_0}-\dfrac{\hbar^2k^2}{2m} \right) \notag\\
&+\dfrac{L}{4\pi}\int_{-\Lambda}^{\Lambda} dk \left({g \sigma_0}+\dfrac{\hbar^2k^2}{2m} \right).
\end{align}
Notice  that the first term is now finite in the limit $\Lambda\to \infty$  while the second term contains linear and cubic divergences in $\Lambda$. Taking the limit $\Lambda\to \infty$ for the first term while keeping $\Lambda$ finite for the second term, \eq{eq:Z0e} becomes
\begin{align}	
-\ln Z_0=&-\dfrac{2L\beta\sqrt{m (g\sigma_0)^3}}{3\pi \hbar} +\dfrac{L\beta\sqrt{m (g\sigma_0)^3}}{\pi \hbar}f_0(\beta g \sigma_0)\notag\\
&+\dfrac{ L\beta g\sigma_0\Lambda }{2\pi}+\dfrac{L\beta\hbar^2\Lambda^3}{12\pi m},
\label{eq:bF0}
\end{align}
where
\be
f_{0}(z)=\int_0^\infty dx\dfrac{1}{z}\ln\left(1-e^{-z\sqrt{x^2+x^4/4}}\right).\label{eq:tf0}
\ee
One has $f_0(z)=-\pi^2/6z^2$ at $z\to\infty$, corresponding to small temperatures in Eq.~(\ref{eq:bF0}). The last two terms  $\propto\Lambda$ and $\propto\Lambda^3$ in \eq{eq:bF0} depend nonuniversally on the cutoff procedure. However they will cancel out with  other contributions in Eq.~(\ref{eq:Om1}) , as we demonstrate below. 

Concerning the third term on the right-hand side of \eq{eq:Om1},  one can show that using the conservation of momentum  $\ac{S_1S_3}=0$  since  $\sigma_{\bf k}$ does not contain $k=0$ mode. 
\begin{widetext}
The remaining terms to be evaluated in \eq{eq:Om1} are
\begin{align}
\dfrac{1}{2}\ac{(S_1)^2S_4}-\dfrac{1}{4}\ac{(S_1)^2(S_3)^2}= \beta\dfrac{G^2m }{\hbar^2}\left[\mathcal F^{(0)}(\ell)+\mathcal F^{(T)}(\ell)+\mathcal F_d(\ell) \right], \label{eq:tbc}
\end{align}
where (see Appendix \ref{app:1L} for the technical details)

\begin{align}
\mathcal F^{(0)}(\ell)=&-\dfrac{1}{2\pi}\left[e^{-2\ell/\xi}(1-2\ell/\xi)+1 \right]-\dfrac{2}{\pi^2}\int_0^\infty dx \cos^2\left(x\ell/\xi\right)\left[\dfrac{2\arctan x}{x(1+x^2)(2+x^2)}+\dfrac{x^2\ln \frac{x}{2+\sqrt{4+x^2}}}{(2+x^2)\sqrt{4+x^2}} \right],\label{eq:f00}\\
\mathcal F^{(T)}(\ell)={}&\dfrac{e^{-2\ell/\xi}}{\pi} \int_0^\infty dx\left[ \dfrac{8e^{-(\ell/\xi)(\sqrt{4+x^2}-2)} \cos(x \ell /\xi)}{x(2+x^2)(4+x^2)}-\dfrac{2}{x\sqrt{4+x^2}} -(2\ell/\xi-1) \dfrac{x}{2\sqrt{4+x^2}}   \right] \widetilde n(x) \notag\\
&+\dfrac{1}{\pi} \int_0^\infty dx\left[ \dfrac{8}{x(2+x^2)(4+x^2)}-\dfrac{4-x^2}{2x\sqrt{4+x^2}}  \right] \widetilde n(x)+ \int_0^\infty dx \dfrac{x^2\sin(2x \ell /\xi)}{\pi(2+x^2)\sqrt{4+x^2}}\widetilde n(x),\label{eq:fT}\\
	\mathcal F_d(\ell)={}&\dfrac{\hbar\Lambda}{4\pi\sqrt{g \sigma_0 m}}\left[1+e^{-2\ell\sqrt{g \sigma_0 m}/\hbar} \left(1-\dfrac{2\ell \sqrt{g \sigma_0 m}}{\hbar} \right)\right].\label{eq:Fd}
\end{align}
\end{widetext}
Here  $\widetilde n(x)=1/( e^{-g \sigma_0\sqrt{x^2+x^4/4}/T}-1)$ and $\xi=\hbar/\sqrt{m g \sigma_0}$. Unlike the contribution obtained in \eq{eq:S1S1}, the latter contributions to the grand canonical potential arise due to the exchange of virtual excitation between the two impurities. We notice that \eq{eq:Fd} is nonuniversal and diverges with $\Lambda$.

Let us now evaluate $S_c^{(1)}$ in Eq.~(\ref{eq:Om1}), which is given by \eq{Sc1}. Regularizing the delta function, one obtains
\begin{align}
{S_c^{(1)}}=-\frac{L\beta g \sigma_0\Lambda}{2\pi}-\frac{  L\beta\hbar^2\Lambda^3}{12\pi m}.
\end{align}
The latter term  cancels the divergent contribution of $-\ln Z_0$ given by the second line of Eq.~(\ref{eq:bF0}). Therefore, formally divergent contribution that arises from the initial Hamiltonian (\ref{eq:h0}) upon using the commutation relations to transform it to the equivalent form (\ref{H}) regularizes the divergence in the one-loop ground state energy of the Lieb-Liniger Bose gas ($G=0$) and no counterterms in the field-theoretical description are needed at this step. However, we do need a counterterm due to the divergent contribution (\ref{eq:Fd}) in a system with impurities ($G\neq 0$), which we now discuss.

As we saw, the quantum field theory  is characterized by ultraviolet divergences. In order to regularize the theory, one has to renormalize its parameters  such that the divergences are absorbed. It is known that power law divergent terms (in the cutoff) depend on a regularization method and do not carry real physics in contrast to logarithmically divergent terms \cite{braaten_quantum_1999,andersen_theory_2004}. In our case, only power law divergences appear.

The one-loop free energy is given by  \eq{eq:F1O1}. In order to absorb the divergence in $\Omega_1(\mu,\bar n_0)$ contained in Eq.~(\ref{eq:Fd}), we need to renormalize the chemical potential. After the substitution $\mu\to\mu+\delta\mu(\Lambda)$ in $F_0(\mu)$ [Eq.~(\ref{eq:F0})] and expansion of the zero-loop free energy in $\delta\mu(\Lambda)$, to  order in $G^0$, we get the one-loop counterterm to be
\begin{align}\label{dG1}
\Delta F_1(\Lambda)={}& -\dfrac{L \mu \delta\mu(\Lambda)}{g}.
\end{align}
In order to remove the linear divergence of Eq.~(\ref{eq:Fd}), one has to impose 
\begin{align}
\delta \mu(\Lambda)=\frac{\Lambda g G^2}{4\pi L\xi_\mu\mu^2}\left[1+e^{-2\ell/\xi_\mu}\left(1-\dfrac{2\ell}{\xi_\mu} \right) \right]\label{eq:dmu},
\end{align}
which renders the one-loop free energy $F_1(\mu)=\Omega_1(\mu,\bar n_0)+\Delta F_1(\Lambda)$ finite.

Now that the one-loop free energy has been regularized,  we split it into the zero temperature term $F_1^{(0)}(\mu)$, and the finite temperature term $F_1^{(T)}(\mu)$. To order $G^2$, we have thus obtained
\begin{subequations}
	\label{F1}
\begin{align}
F_1(\mu)=F_1^{(0)}(\mu)+F_1^{(T)}(\mu),
\end{align}
\begin{align}
F_1^{(0)}(\mu)=&-\dfrac{2 L \mu}{3\pi\xi_\mu }+\dfrac{2 G}{\pi\xi_\mu}+\dfrac{G^2m}{\hbar^2}\mathcal F^{(0)}(\ell)\notag\\
&-\dfrac{G^2m}{2\pi\hbar^2}\left[ 1+e^{-2\ell/\xi_\mu}\left(1-\dfrac{2\ell}{\xi_\mu} \right)\right],  \label{eq:F10}
\end{align}
\begin{align}
F_1^{(T)}(\mu)={}&\dfrac{ L \mu}{\pi\xi_\mu }f_0(\beta\mu)+\dfrac{G}{\pi\xi_\mu}f_1(\beta \mu)+\dfrac{G^2m}{\hbar^2}\mathcal F^{(T)}(\ell)\notag\\
&-\dfrac{G^2m}{4\pi\hbar^2}\left[1+e^{-2\ell/\xi_\mu }\left(1 -\dfrac{2\ell}{\xi_\mu}\right) \right]f_1(\beta\mu).\label{eq:F1T}
\end{align}
\end{subequations}
Here we introduced
\be
f_1(z)=-2zf_0'(z)-3f_0(z).\label{eq:tf1}
\ee
At large $z$, corresponding to small temperatures, both $f_0(z)$ and $f_1(z)$ behaves as $-\pi^2/6z^2$. We recall that functions $\mathcal F^{(0)}(\ell)$ and $\mathcal F^{(T)}(\ell)$ are given by Eqs.~(\ref{eq:f00}) and (\ref{eq:fT}) respectively, as well as the definition (\ref{ximu}). Therefore, the Landau free energy of the system to one-loop order is given by $F_0(\mu)+F_1(\mu)$ where the corresponding contributions are given by Eqs.~(\ref{eq:F0}) and (\ref{F1}).

\section{Helmholtz free energy}
\label{sec5}

In this section we use the Landau free energy, which is a function of the chemical potential of the system, to calculate the Helmholtz free energy, which depends on the system density. The latter quantity will enable us to find the effective interaction between the two impurities in the following section.

The Helmholtz free energy is defined via Legendre transformation of the Landau free energy 
\be
A(\bar n,\beta,\ell)=F(\mu,\beta,\ell)+\mu \bar nL.\label{eq:en}
\ee
Here, $\bar nL=N$ is the number of particles. We point out that in \eq{eq:en}, the chemical potential on the right hand side of the equation   should be considered as a function of the density, i.e., $\mu\equiv \mu(\bar n)$. 
The expression of $\mu(\bar n)$ can be obtained by inverting $\bar n(\mu)$ given by Eq.~(\ref{eq:density}).
In our case, the Landau free energy is given to one-loop order as $F=F_0+F_1^{(0)}+ F_1^{(T)}$ where these three terms are given  to order $G^2$ by Eqs.~(\ref{eq:F0}), (\ref{eq:F10}) and (\ref{eq:F1T}) respectively. 
Inverting \eq{eq:density} leads to the chemical potential of the system
\begin{align}
\mu(\bar n)={}&\mu_0\biggl\{1+\dfrac{2G}{L\mu_0}-\dfrac{G^2}{2L \xi\mu_0^2}\left[1+e^{-2\ell/\xi}\left(1-\dfrac{2\ell}{\xi} \right)\right]\notag\\
&-\dfrac{\sqrt{\gamma}}{\pi}\left[1+\dfrac{f_1(\beta \mu_0)}{2} \right]+\dfrac{\sqrt{\gamma}G^2}{L\xi\mu_0^2}\bar n\notag\\
&\times\partial_{\sigma_0}\left.\left[\mathcal F^{(0)}(\ell)+ \mathcal F^{(T)}(\ell)\right]\right|_{\sigma_0=\bar n}+\mathcal O\big( G^3) \biggr\},\label{eq:mun}
\end{align}
where $\mu_0=g \bar n$. In \eq{eq:mun}, 
\be
\xi=1/\bar n\sqrt{\gamma}
\ee
denotes the healing length of the weakly-interacting Bose gas. Performing the Legendre transformation (\ref{eq:en}), the Helmholtz free energy reads  to one-loop (i.e., to order $\sqrt{\gamma}$ that denotes the first quantum correction) and to order $G^2$ 
\begin{align}
A(\bar n,\beta,\ell)={}
&\dfrac{\hbar^2\bar n^2 N\gamma}{2m}\biggl\{1-\dfrac{4\sqrt{\gamma}}{3\pi}\left[1-\frac{3}{2}f_0\left(\dfrac{\hbar^2\bar n^2\beta\gamma}{m}\right)\right]\notag\\
&+\mathcal O(\gamma)\biggr\}+2G\bar n\left[1+\mathcal O(\gamma)\right]\notag\\
&-\dfrac{G^2m}{\hbar^2\sqrt{\gamma}}\biggl\{1+e^{-2\ell/\xi}\notag\\& -{\sqrt{\gamma}}\left[\mathcal F^{(0)}(\ell)+\mathcal F^{(T)}(\ell)\right]+\mathcal O(\gamma)\biggr\}\notag\\
&+\mathcal O(G^3).\label{eq:enf}
\end{align}
In the previous expression, $\sigma_0$ should be replaced by $\bar n$ in  expressions for the spectrum $\varepsilon_k$ that enter $\mathcal F^{(0)}$ and $\mathcal F^{(T)}$. At $G=0$, \eq{eq:enf} describes the one-loop free energy of the Lieb-Liniger Bose gas, as we discuss in Appendix \ref{app:G0}. 

\section{Effective interaction between impurities}

\label{sec:Eff}

In this section we calculate the effective interaction between impurities $U(\ell,T)$ and obtain an analytic expression valid at arbitrary impurity separations $\ell$. We consider both cases of zero temperature and finite temperature. The expression for $U(\ell,T)$ will be given in terms of a single integral of an elementary function. The limiting cases are evaluated and discussed.

The effective interaction between two-impurities at separation $\ell$ is extracted from the Helmholtz free energy \eqref{eq:enf} as 
\be
U(\ell,T)=A(\bar n,\beta=1/T,\ell)-A(\bar n,\beta=1/T,\ell\to\infty).\label{eq:Uint}
\ee
Here we subtract the energy of two isolated impurities at $\ell\to\infty$ in order to obtain the interaction term. Equation (\ref{eq:Uint}) can be split into a zero and finite temperature contribution,
\be
U(\ell,T)=U^{(0)}(\ell)+U^{(T)}(\ell).\label{111}
\ee 
The first term $U^{(0)}(\ell)$ describes the interaction at zero temperature and originates from $\mathcal{F}^{(0)}(\ell)$ of Eq.~(\ref{eq:enf}), while the second term  $U^{(T)}(\ell)$ describes the effect of thermal fluctuations on the interaction and arises from  $\mathcal{F}^{(T)}(\ell)$ of Eq.~(\ref{eq:enf}).

\subsection{Zero temperature case}

Substituting \eq{eq:enf} into Eq.~(\ref{eq:Uint}) at $T=0$, one obtains the zero temperature interaction between two-impurities. At lowest nontrivial order (which is $G^2$) it takes the form
\begin{align}
U^{(0)}(\ell)=	&-\dfrac{G^2 m}{\hbar^2 \sqrt{\gamma}}e^{-2\ell/\xi}\left\{1+\dfrac{\sqrt{\gamma}}{2\pi}\left[ 1-\dfrac{2\ell}{\xi}+J\left(\dfrac{2\ell}{\xi}\right)\right]\right\}\notag\\
&-\dfrac{G^2 m}{\pi^2\hbar^2 }\int_0^\infty dx\dfrac{x^2\ln\left(\frac{x}{2+\sqrt{4+x^2}}\right)\cos\left(\frac{2x \ell}{\xi}\right)}{(2+x^2)\sqrt{4+x^2}}.
\label{eq:U0}
\end{align}
The function $J(z)=\frac{4e^z}{\pi}\int_0^\infty dx \frac{\arctan x \cos(xz)}{x(1+x^2)(2+x^2)}$ in \eq{eq:U0} can be expressed as
\begin{align}
J(z)=&-\dfrac{e^{z(1+\sqrt{2})}}{2}\text{Ei}(-z\sqrt{2}-z) -\dfrac{e^{z(1-\sqrt{2})}}{2}\text{Ei}(z\sqrt{2}-z)  \notag\\ 
&+e^{2z}\text{Ei}(-2z)-e^{z}\text{Ei}(-z)-e^{z(1-\sqrt{2})}\ln(1+\sqrt{2})\notag\\
&+\ln (2z)+\gamma_E,
\end{align}
where $\gamma_E\approx 0.5772$ is the Euler constant, while the exponential integral is defined as $\text{Ei}(z)=-\int_{-z}^\infty dt \frac{e^{-t}}{t}$. At small and large arguments, $J(z)$ has the asymptotic behavior
\begin{align}
J(z)=\begin{cases}
\ln(4\sqrt{2}-4)+\ln(4\sqrt{2}-4) z+\mathcal{O}(z^2),\\
\ln(2z)+\gamma_E-\frac{1}{2z}+\mathcal{O}\left(\frac{1}{z^2}\right),
\end{cases}
\end{align}
while in the crossover region it monotonously increases. 

For small values of $\gamma$ (i.e., where our perturbation theory applies), the effective interaction (\ref{eq:U0}) is attractive. One can distinguish two regimes arising at different impurity separations. At short separations $\ell\lesssim \xi$, the interaction decays exponentially according to the leading, zero-loop, first term of \eq{eq:U0}, 
\be
U^{(0)}(\ell)=-\frac{G^2m}{\hbar^2\sqrt{\gamma}}e^{-2\ell/\xi}. \label{Ucl}
\ee 
This result is in accordance with various previous studies \cite{klein_interaction_2005,recati_casimir_2005,dehkharghani_coalescence_2018}.

The form (\ref{eq:U0}) contains the first quantum correction to the result  (\ref{Ucl}) arising from one-loop contribution. It is smaller by a factor $\sqrt{\gamma}$ from the classical contribution (\ref{Ucl}), however, at large distances the classical result is exponentially small and the quantum correction becomes the dominant one. Therefore, the second regime occurs at large distances where an algebraic decay of the quantum contribution prevails. The effective interaction law can be determined in the following way. At large distances, $\ell \gg \xi$, we expand the last term in \eq{eq:U0}, while keeping $x\ell/\xi\sim1$. The large distance behavior is dictated by the logarithmic singularity of the integrand at small $x$. One finds
\begin{align}
U^{(0)}(\ell)=&-\dfrac{G^2m}{\pi^2\hbar^2}\int_0^\infty dx \dfrac{x^2\cos \left( {2x \ell}/{\xi}\right)\ln x }{(2+x^2)\sqrt{4+x^2}}\notag\\
=&-\dfrac{G^2 m}{32\pi \hbar^2}\dfrac{\xi^3}{l^3}\left[1+\dfrac{15}{8}\dfrac{\xi^2}{l^2}+\mathcal O\big( \xi^4/\ell^4\big) \right]. \label{eq:int1}
\end{align}
The last expression shows that the induced inter-impurity interaction decays as a cubic power law of the inverse distance, in accordance with previous studies \cite{schecter_phonon-mediated_2014,reichert_casimir_2018}. We point out that even though only the linear part of the Bogoliubov dispersion relation is needed to obtain the leading term of \eq{eq:int1}, the subleading term $\propto 1/\ell^5$ requires the leading correction to the linear dispersion, which is  cubic in the momentum, i.e,  $\vep_k\approx (\hbar^2/\xi m) |k|+(\hbar^2\xi/8m)|k|^3$. However, it is known that this spectrum is not valid when $k\to 0$ \cite{pustilnik_low-energy_2014,imambekov_one-dimensional_2012}. Indeed, below a momentum $k^*\sim \gamma^{1/4}/\xi$, the leading correction to the linear dispersion relation of quasiparticles in the one-dimensional quantum liquid is quadratic rather than cubic and the dispersion relation becomes  $\vep_k\approx (\hbar^2/\xi m) |k|+\hbar^2k^2/2m^*$ where $m/m^*\sim \gamma^{1/4}$. This implies that the subleading correction in \eq{eq:int1} is valid at distances $\ell \ll \xi/\gamma^{1/4}$. This is still huge with respect to $\xi$, since at weak interaction $\gamma\ll 1$.

One can  define a crossover distance $\ell_c^{(0)}$ at which the behavior of the induced impurity interaction changes from the exponential (\ref{Ucl}) to the algebraic decay (\ref{eq:int1}). The crossover distance is obtained by solving  $\exp({-2\ell_c^{(0)}/\xi})=\xi^3 \sqrt{\gamma}/32\pi\left(\ell_c^{(0)}\right)^3$, yielding
\be
\ell_c^{(0)}\approx\dfrac{\xi}{2}\ln (32\pi/\sqrt{\gamma}).
\label{eq:L0T0}
\ee
This equation shows that the crossover scale depends weakly (logarithmically) on the interaction. Therefore, at distances $\ell\lesssim \ell_c^{(0)}$, the impurity interaction is given by Eq.~(\ref{Ucl}), while at large separations $\ell\gg\ell_c^{(0)}$, it takes the form (\ref{eq:int1}). At intermediate distances, where $\ell$ is of the order of $\ell_c^{(0)}$, one needs the full expression (\ref{eq:U0}) to characterize the interaction.

The interaction between impurities at zero temperature was recently calculated in Ref.~\cite{reichert_casimir_2018} using a complementary approach based on the Gross-Pitaevskii equation, see Appendix \ref{app:expre}. It leads to the expression (\ref{eq:U0b}) that is seemingly different from Eq.~(\ref{eq:U0}). We have, however, shown that the two expressions are identical, as it must be the case.

\subsection{Finite temperature case}

Let us now consider the effects of finite temperature on the induced interaction between impurities. We first notice that at high temperature the density fluctuations of a Bose gas become large and thus our starting point, in particular the assumption (\ref{n}), is invalid. However, at $T\ll \hbar^2 \bar n^2\sqrt\gamma/m=\mu_0/\sqrt{\gamma}$, a weakly-interacting one-dimensional Bose gas is characterized by small density fluctuations and remains in the quantum coherent quasi-condensate regime \cite{kheruntsyan_pair_2003}, which we consider in the following.

\begin{figure}
	\centering
	\includegraphics[width=0.9\columnwidth]{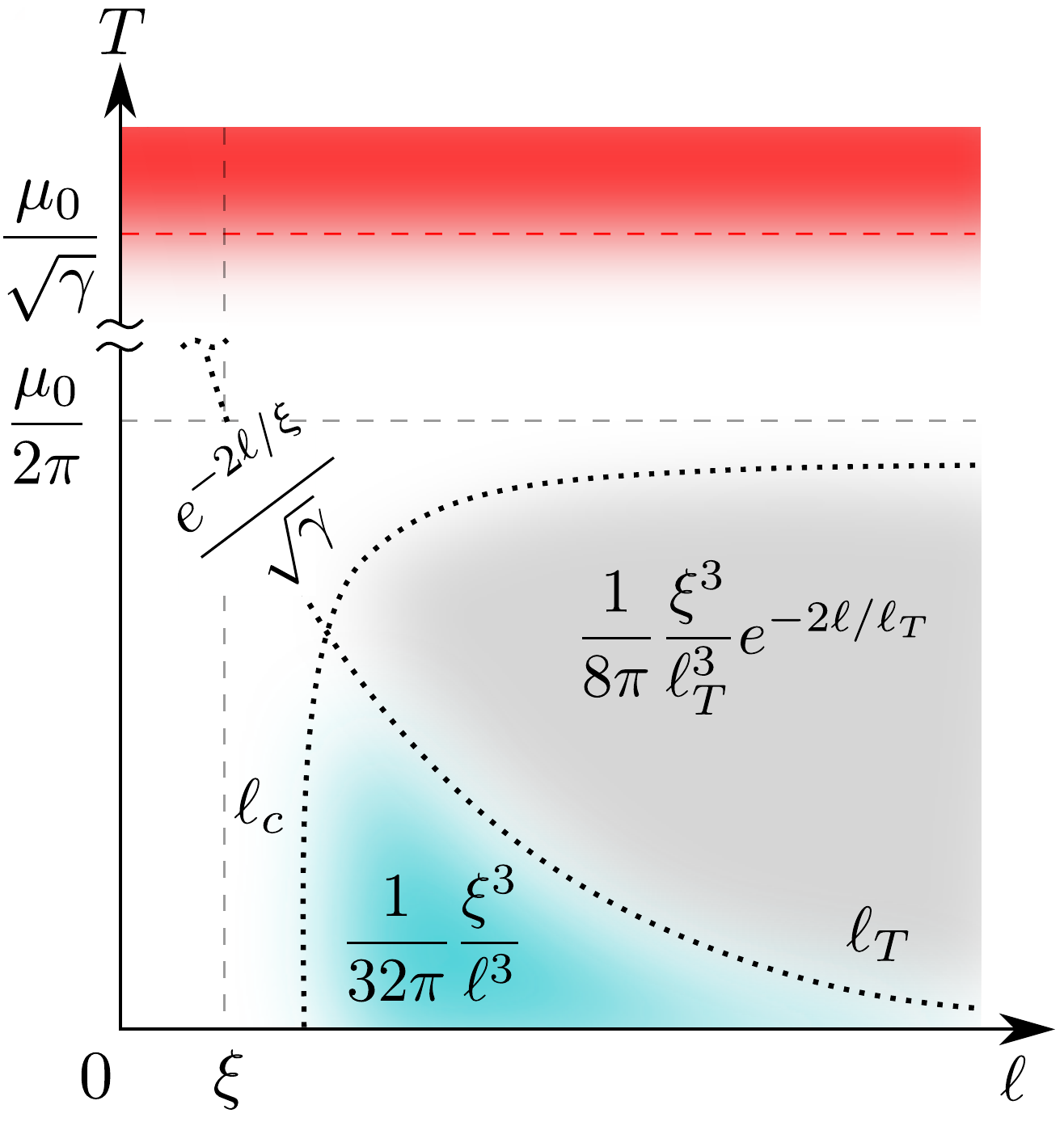}
	\caption{Schematic diagram representing the dominant behavior of the induced interaction $-\frac{\hbar^2}{G^2m} U(\ell,T)$ as a function of the distance between the impurities $\ell$ and the temperature $T$. Here $\mu_0=\hbar^2 \bar n^2\gamma/m$, $\xi=1/\bar n\sqrt{\gamma}$, and $\gamma=mg /\hbar^2\bar n$. The crossover distance $\ell_c$ (dotted line) is defined by Eq.~(\ref{eq:L0T}), while the thermal length $\ell_T$ (dotted line) is given by Eq.~(\ref{lT}). The different regimes of $U(\ell,T)$ are: (i) the exponential decrease as $\exp(-2\ell/\xi)$ for $\ell\ll\ell_c$ (white), (ii) the power law $1/\ell^3$ decay for $\ell_c\ll\ell\ll\ell_T$ (blue), (iii) the exponential decrease as  $\exp({-2\ell/\ell_T})$ for $\ell\gg\text{max}(\ell_c,\ell_T)$ (gray). The high temperature region $T\gg\mu_0/\sqrt{\gamma}$ where the perturbation theory breaks down is represented in red and is not considered in this work.}
	\label{fig:2d}
\end{figure}

Already at the qualitative level we can distinguish different regimes of the induced interaction. Finite temperature introduces the thermal length
\begin{align}\label{lT}
\ell_T=\frac{\hbar v}{2\pi T},
\end{align}
where $v$ is the sound velocity (see Appendix \ref{app:G0}). At distances longer than the thermal length, the quasi-long range order of the Bose gas is lost and the correlations decay exponentially. We thus expect that the induced interaction behaves in a similar way. At small temperature, $\ell_T$ is large. The induced long-range interaction is thus practically not affected by the temperature as long as the impurity separation $\ell$ is smaller than $\ell_T$. By increasing the temperature $\ell_T$ decreases and modifies the long-range interaction at $\ell\gtrsim\ell_T$. The schematic diagram representing the characteristic behavior of the induced interaction and the boundaries between different regimes is shown in Fig.~\ref{fig:2d}.

Using Eqs.~(\ref{eq:enf}) and (\ref{eq:Uint}),  the finite temperature contribution to the induced interaction (\ref{eq:U0}) is given by
\begin{align}
U^{(T)}(\ell)={}&\dfrac{G^2m}{\hbar^2}\left[\vphantom{\dfrac{1}{1}}\mathcal{F}^{(T)}(\ell)-\mathcal{F}^{(T)}(\ell)|_{\ell\to\infty}\right]. \label{ut}
\end{align}
Here the function $\mathcal F^{(T)}(\ell)$ is defined by \eq{eq:fT}.  The induced interaction (\ref{111}) at finite temperature we express as
 \begin{align}
U(\ell,T)=-\frac{G^2m}{\hbar^2\sqrt{\gamma}}e^{-2\ell/\xi}+U_{\mathrm{f}}(\ell,T).\label{eq:UtotT}
\end{align}
The first term on the right hand side is the classical result  (\ref{Ucl}), while the second one represents the fluctuation contribution. Using the equivalent expression for $U^{(0)}(\ell)$ of \eq{eq:U0} that is given in Appendix~\ref{app:expre}, we express the fluctuation contribution as 
	\begin{align}
		U_{\mathrm{f}}(\ell,T)={}& \dfrac{G^2m}{ 2\pi\hbar^2}e^{-2\ell/\xi}  f(\ell/\xi,T)+U_{\mathrm{f{}l}}(\ell,T).  \label{Uff}
	\end{align}
Here the fluctuation correction of the exponential interaction is encoded in the function  
\begin{align}
f(z,T)={}&\int_0^\infty dx \biggl[ \dfrac{8e^{-z(\sqrt{4+x^2}-2)}\cos (zx)}{x(2+x^2)(4+x^2)}-\dfrac{2}{x\sqrt{4+x^2}}\biggr]\notag\\
&\times\left[1+2\wt{n}_b(x,T)\right]\notag\\
&+(2z-1)\left[1-\int_0^\infty dx\dfrac{x }{\sqrt{4+x^2}}\widetilde n_b(x,T)  \right], \label{wtK}
\end{align}
where the occupation factor is
\begin{align} 
\wt{n}_b(x,T)=\dfrac{1}{e^{(\mu_0/T) \sqrt{x^2+x^4/4}}-1}.
\end{align}
We recall the definition $\mu_0=g\bar n$. At zero-temperature, $f(z,T)$ describes the quantum correction to the exponential part of the interaction given by the first line of Eq.~(\ref{eq:U0}). The temperature effect on the long-range part of the interaction in Eq.~(\ref{Uff}) is given by
\begin{align}
U_{\mathrm{f{}l}}(\ell,T)={}&\dfrac{G^2m}{ 2\pi\hbar^2}\int_0^\infty  \dfrac{dx\, x^2\sin(2\ell x/\xi)}{(2+x^2)\sqrt{4+x^2}}\left[1+2\wt{n}_b(x,T) \right].\label{2nd}
\end{align}
Let us analyze the case of distances $\ell$ longer than $\xi$.
A numerical comparison of the two terms in Eq.~(\ref{Uff}) shows that $U_{\mathrm{f}}(\ell,T)\approx U_{\mathrm{f{}l}}(\ell,T)$ at temperatures below $\mu_0/2\pi$. We notice that in that case $\ell_T> \xi$. At temperatures above $\mu_0/2\pi$ (and below $\mu_0/\sqrt{\gamma}$), the fluctuation induced part of the interaction $U_{\mathrm{f}}(\ell,T)$ is determined by the first term in Eq.~(\ref{Uff}).

Consider the lowest temperatures and distances longer than $\xi$. The dominant term in the interaction is the second one of Eq.~(\ref{Uff}) given by Eq.~(\ref{2nd}). We first notice that the Bose-Einstein distribution weights the integrand in such way that the region  $x\lesssim T/\mu_0=\xi/2\pi \ell_T$ gives the main contribution. In addition, if $\xi/2\pi \ell_T$ is smaller than one, one can safely linearize the Bogoliubov dispersion and then express the Bose-Einstein distribution function as a series
\be
\wt n_b(x,T)=\sum_{j=1}^\infty e^{-2\pi j (\ell_T/\xi)  x}. \label{nb2}
\ee
Using \eq{nb2}, and expanding the remaining part of the integrand for $\ell\gg \xi$, one can perform the integration over $x$ and  then resum  the series to obtain the leading contribution at large distances. This yields
\begin{align}
U_{\mathrm{f}}(\ell,T)\approx U_{\mathrm{f{}l}}(\ell,T)=-\dfrac{G^2m}{32\pi\hbar^2}\dfrac{\xi^3}{\ell_T^3} \dfrac{\cosh (\ell/\ell_T)}{\sinh^3(\ell/\ell_T)}.\label{eq:ft1}
\end{align}
At zero temperature $\ell_T$ diverges and the induced interaction takes the form 
\be
U_{\mathrm{f{}l}}(\ell,T)=-\dfrac{G^2m}{32\pi\hbar^2} \dfrac{\xi^3}{\ell^3},
\label{eq:ft2}
\ee
in agreement with Eq.~(\ref{eq:int1}). At  high temperatures we have $\ell_T\ll\ell$. Since the quasi-long-range order in a one-dimensional system of bosons is lost at distances beyond $\ell_T$, one  expects the absence of the long-range interaction between impurities. This is indeed true as
\eq{eq:ft1} becomes \cite{schecter_phonon-mediated_2014} 
\begin{align}
U_{\mathrm{f{}l}}(\ell,T)=-\dfrac{G^2m}{8\pi\hbar^2}\dfrac{\xi^3}{\ell_T^3}e^{-2\ell/\ell_T}. \label{eq:UexpT1}
\end{align}
In the regime of intermediate temperatures ($\ell_T\sim \ell$) the induced interaction is given by Eq.~(\ref{eq:ft1}). 

The total interaction at $T$ smaller than $\mu_0/2\pi$ is obtained when the classical part (\ref{Ucl}) is added to $U_{\mathrm{f{}l}}(\ell,T)$. A comparison of the two contributions enables us to define the crossover distance $\ell_c$ below which the classical result dominates. At high temperatures ($\ell_T<\ell_c^{(0)}$)  the crossover scale can be approximated by
\be
\ell_c\approx\frac{\xi}{2} \dfrac{\ell_T}{\ell_T-\xi}\ln \left(\frac{8\pi \ell_T^3}{\xi^3\sqrt{\gamma}}\right).
\label{eq:L0T}
\ee
At low temperatures ($\ell_c^{(0)}<\ell_T$) the crossover scale is the one of the zero-temperature case given by Eq.~(\ref{eq:L0T0}): $\ell_c\equiv\ell_c^{(0)}$. The crossover distance $\ell_c$ is represented by the dotted line in Fig.~\ref{fig:2d}. 

At $T\gtrsim\mu_0/2\pi$ we need to evaluate the first term in \eq{Uff}. We can do it analytically in the case of large distances. Consider $f(z,T)$ of Eq.~(\ref{wtK}). The second integral in the expression (\ref{wtK}) is a positive monotonic function of temperature that behaves as $\pi^2T^2/12\mu_0^2$ at $T\gg \mu_0$ and as $\pi T/2\mu_0$ in the opposite limit $T\ll \mu_0$. The temperature-independent part in the first integral is also monotonic that decreases from $-\ln 4$ at $z=0$ to $-\ln z$ at large $z$. The temperature dependent part in the same integral is again monotonic and behaves as $-\pi T z/\mu_0+\ln z$ at large $z$. Interestingly, the logarithmic term cancels in the sum and we obtain
\begin{align}
f(z,T)=\left\{2-\frac{\pi T}{\mu_0}[1+h(T/\mu_0)]\right\}z+\mathcal{O}(z^0).
\label{f}
\end{align}
Here $h(0)=1$ and $h(y\to \infty)=\pi/6y$. Therefore $f(z,T)$ changes the sign at some temperature, which we numerically find to be $T^*\approx 0.526\mu_0$. At $T<T^*$ we obtained  $f(z,T)>0$, while $f(z,T)<0$ at $T>T^*$. We recall that our approach is valid at $T\ll\mu_0/\sqrt{\gamma}$. Since $\gamma\ll 1$ at weak interaction, $T^*$ is within this region. We thus obtain the induced interaction at $T\gtrsim\mu_0/2\pi$ and at distances longer than $\xi$ to be
\begin{align}
U(\ell,T)=-\frac{G^2m}{\hbar^2\sqrt{\gamma}}e^{-2\ell/\xi}\left[1-\frac{\sqrt{\gamma}}{2\pi}f(\ell/\xi,T)\right].\label{eqU}
\end{align}
Equation (\ref{eqU}) shows how the prefactor of the classical exponential term gets modified by the fluctuations. It resembles the first line of Eq.~(\ref{eq:U0}). However, in Eq.~(\ref{eq:U0}) that was not particularly important  due to the existence of the long-range part. In the present case, the long-range part is suppressed by thermal fluctuations and the modification of the prefactor can, in principle, play an important role. The correction term $f(\ell/\xi,T)$ in Eq.~(\ref{eqU}) can even change the sign of the interaction as a function of the distance and the temperature, since $f(\ell/\xi,T)$ is positive at $T<T^*$ and grows linearly with $\ell$. However, the exponential term makes the interaction very small when it changes the sign. It would be thus interesting to explore the same question at strong coupling where the change of sign might occur at smaller $\ell$.

We finally study the induced interaction (\ref{eq:UtotT}) at small distances $\ell\ll\xi$. We find the leading low temperature corrections
\begin{gather}
f(\ell/\xi,T)=-\ln 4-1-\frac{\pi^2 T^2}{8\mu_0^2},\\
U_{\mathrm{f{}l}}(\ell,T)=\frac{G^2m}{2\pi\hbar^2}\left(\frac{\pi}{2}+\frac{\pi^4T^4}{15 \mu_0^4}\frac{\ell}{\xi}\right).
\end{gather}
We notice that $U_{\mathrm{f{}l}}(\ell,T)$ at small distances tends to a temperature-independent constant, while $f(\ell/\xi,T)$ decreases with the temperature. We therefore obtain
\begin{align}
U(\ell\to 0,T)=-\frac{G^2m}{\hbar^2\sqrt{\gamma}}-\frac{G^2m}{2\pi\hbar^2}\left(\ln 4+1-\frac{\pi}{2}+\frac{\pi^2 T^2}{8\mu_0^2}\right)
\label{u0}
\end{align}
at $T\ll \mu_0$. In the opposite limit $T\gg \mu_0$, we numerically obtained that the expression in parentheses in Eq.~(\ref{u0}) changes and becomes proportional to $\sqrt{T/\mu_0}$. At the border of applicability where the temperature is of the order $\mu_0/\sqrt{\gamma}$, the fluctuation correction scales as $\gamma^{-1/4}$ and thus it is smaller than the classical term that scales as $\gamma^{-1/2}$.

\section{Conclusions and discussions}
\label{sec:Conclusion}

In this work we studied the effective Casimir-like interaction between two static impurities immersed in a weakly interacting one-dimensional Bose gas. We used a field-theoretical approach and developed a microscopic theory, which enabled us to calculate the effective interaction between impurities mediated by quasiparticles. At weak interaction they are described by the Bogoliubov spectrum, which is linear at low and quadratic at high momenta. Due to weak anharmonic terms in the Hamiltonian, Bogoliubov quasiparticles experience a residual interaction which is important in a consistent description of the Casimir-like interaction. Since we described the quasiparticles (practically) at all momenta, we were able to obtain the expression for the induced interaction valid at arbitrary distances between the impurities. At zero temperature, our results show that the problem does not contain intermediate regimes and that there is just one crossover scale in the problem that is the healing length $\xi$ (up to a weak logarithmic dependence on the interaction).

On the technical level, from the Helmholtz free energy, we evaluated the mean-field (zero-loop) interaction and its leading quantum correction (one-loop) that at zero temperature are given by Eq.~(\ref{eq:U0}). The former contribution is responsible for the short-range interaction decaying exponentially at distances longer than the healing length $\xi$ [see Eq.~(\ref{Ucl})]. At zero temperature, the leading quantum correction to the induced interaction becomes a dominant contribution since it decays algebraically at distances longer than $\xi$. We notice that in order to  calculate the quantum correction the residual interaction between quasiparticles has to be taken into account \cite{schecter_phonon-mediated_2014}. In this paper we confirmed our recent zero temperature results obtained in Ref.~\cite{reichert_casimir_2018} using a complementary approach. Here we also studied the effect of finite temperature on the effective interaction and obtained analytical expressions at all impurity distances [Eqs.~\eqref{eq:UtotT}--\eqref{2nd}]. The different regimes of the induced interaction are shown in Fig.~\ref{fig:2d}. As a side result, from our approach we were able to obtain the expressions for several thermodynamic quantities of the Lieb-Liniger model at weak interaction (see Appendix~\ref{app:G0}), which agree with well established results for that integrable model.

In our treatment we studied the impurities that are weakly coupled to the Bose liquid, thus creating a small disturbance of the system density around impurity positions. The scaling $\propto 1/\ell^3$ of the induced interaction at large distances is, however, much more robust and also exists for an arbitrary coupling strength of impurities to the Bose gas \cite{schecter_phonon-mediated_2014}. However, unlike the phenomenological method of Ref.~\cite{schecter_phonon-mediated_2014} that can study the induced interaction only at large distances $\ell\gg\xi$, our work does not have such limitation and covers all $\ell$, but is limited to the weak interaction between bosons. It might be possible to extend our study to the case of arbitrary coupling $G$ between bosons and the impurities. Namely, the Hamiltonian of the impurity is linear with respect to the boson density and in principle it could be diagonalized from the beginning, and not treated as a perturbation. We also notice that the limit $G=\infty$ is singular and thus could lead to a different scaling of the long-range interaction \cite{recati_casimir_2005,yu_casimir_2009}. If so, the crossover between large but finite $G$ and the case of infinite $G$ is another interesting problem. We leave these questions for future studies.

Throughout the paper we studied the case of two impurities that interact with the Bose gas with equal couplings. In a more general case of different coupling constants $G_1$ and $G_2$, which are both weak, the final result for the induced interaction has still the form Eq.~(\ref{eq:U0}) where $G^2$ should be replaced by $G_1 G_2$. At finite temperature one should use the same replacement in Eqs.~\eqref{eq:UtotT}--\eqref{2nd}.

Let us comment on different results  for the present problem and the methods used in the earlier works. In Refs.~\cite{klein_interaction_2005,recati_casimir_2005,dehkharghani_coalescence_2018} the calculation of the induced interaction was based on the mean-field Gross-Pitaevskii approach and leads to the classical result (\ref{Ucl}). On the contrary, in Ref.~\cite{schecter_phonon-mediated_2014} is used a phenomenological low-energy mobile impurity approach within the Luttinger liquid theory which leads to $1/\ell^3$ result (\ref{eq:int1}) at long distances but a delta function in the opposite limit. This is not surprising, since the quasiparticles of the liquid in Ref.~\cite{schecter_phonon-mediated_2014} are well described only at low momenta when their spectrum is linear. On the contrary, the approaches of  Refs.~\cite{klein_interaction_2005,recati_casimir_2005,dehkharghani_coalescence_2018} describe well the quasiparticles at all momenta by the Bogoliubov spectrum (\ref{eq:Bogspectrum}), but treat the problem on the mean-field level and thus leads to the exponential interaction. In our recent work \cite{reichert_casimir_2018} we were able to account for the effect of quantum fluctuations within the Gross-Pitaevskii formalism, leading to the full crossover behavior of the induced interaction (see Appendix \ref{app:expre}) that contains both limits of short [Eq.~(\ref{Ucl})] and long impurity separation [Eq.~(\ref{eq:int1})]. Finally, in this work we developed a field-theoretical approach that confirms the zero-temperature expression of Ref.~\cite{reichert_casimir_2018} and gives the results at finite temperature. The zero-loop order in the present work corresponds to the mean-field exponential interaction, while one-loop order is necessary to obtain the result at long distances. One-loop order is thus necessary to describe the effect of quantum fluctuations.

We now discuss experimental realizations where our results could be tested. The role of impurities is typically played by neutral atoms of a different specie than the ones forming the Bose gas. Such neutral atoms do interact via another induced interaction of dipole-dipole type, which is of the electromagnetic origin and known as London-van der Waals or Casimir-Polder interaction, depending on the distances. The former decays as $\propto 1/\ell^6$ crossing over into $-\mathcal C\hbar c\alpha_p^2/\pi \ell^7$ at large distances \cite{casimir_influence_1948}. Here $\mathcal C$ is a number of order unity, while $\alpha_p$ is the static polarisability of the impurities. By comparing our result (\ref{eq:int1}) and the Casimir-Polder one, we  find the distance
\begin{align}
\ell_0/\xi\approx (32 \mathcal C)^{1/4}\sqrt{\gamma\alpha_p n^3  g \sqrt{c}/G \sqrt{v}}
\end{align}
where the two interactions are equal. Here $v$ is the sound velocity and $c$ is the speed of light. In a setup with Yb atoms as impurities, the polarisability is $\alpha_p\approx 21$ \AA$^3$. Using the typical values   for a $^{87}\mathrm{Rb}$ Bose gas \cite{hofferberth_probing_2008}: $\gamma=0.005$ and $n=60\, \mu \mathrm{m}^{-1} $, we obtain $v=0.32\,\mathrm{cm/s}$ and $\xi=0.24\,\mathrm{\mu m}$. For $G=4g$ we find $\ell_0\approx 0.1 \xi$. Thus at distances longer than $\xi/10$, the Casimir-like interaction is the dominant one in the above example. For the two impurities at separation $\xi$, Eq.~(\ref{eq:U0}) gives the experimentally measurable value of $0.3\,\textrm{kHz}$, while the Casimir-Polder interaction has six orders of magnitude smaller value. 

Finally, we notice that the attraction of two identical impurities (\ref{eq:U0}) could cause the formation of their quantum-mechanical bound state known as a bipolaron \cite{roberts_impurity_2009,casteels_bipolarons_2013,camacho-guardian_bipolarons_2018}. To reach that state,  the potential energy should exceed the kinetic energy of the relative particle. At separation $\ell_b$ between impurities, we estimate the typical kinetic energy of the relative particle to be $\hbar^2/M\ell_b^2$, where $M$ is the impurity mass. This leads to the condition
\begin{align}\label{eq:bip}
\frac{\hbar^2}{M\ell_b^2}<\frac{G^2 m}{\hbar^2\sqrt{\gamma}}e^{-2\ell_b/\xi},
\end{align}
where we assumed $\ell_b\lesssim\xi$. Equation (\ref{eq:bip}) shows that large $G$ makes the bipolaron radius smaller. At the border of applicability of our theory, $G=g/\sqrt{\gamma}$, we find the bipolaron radius to be $ \xi\sqrt{m/M}\gamma^{1/4}$ that is smaller than $\xi$ at weak interaction and not too light impurities. Our estimates neglect other interactions between impurities, such as van der Waals one, which might play some role. A more complete theory of formation of bipolarons in one-dimensional Bose gases  requires a separate detailed study.

\begin{acknowledgments}
This study has been partially supported through the EUR grant NanoX ANR-17-EURE-0009 in the framework of the ``Programme des Investissements d’Avenir".
	\end{acknowledgments}

\appendix
\begin{widetext}
\section{Calculation of the one-loop terms}
\label{app:1L}

In this appendix we give technical details of the calculation of the terms of \eq{eq:tbc}. The first term is given by
\begin{align}
\dfrac{1}{2}\ac{( S_1)^2  S_4}={}&\dfrac{1}{2}\sum_{\vk{1},..,\vk{6}}\Gi{1}\Gi{2}\Gf{3}{4}{5}{6}\ac{\Pi_{i=1}^6\sigma_{\vk{i}}}\delta_{\om_1,0}\delta_{\om_2,0}\delta_{\vk{3}+\vk{4}+\vk{5}+\vk{6},0}\notag\\
={}&\beta\dfrac{G^2\hbar^8}{4m^4}\int \dfrac{dk}{2\pi}\cos^2\left(\dfrac{k\ell}{2}\right)\dfrac{k^4}{\vep_k^4}\int_{|k'|<\Lambda} \dfrac{dk'}{2\pi}\dfrac{k'^2(k^2+k'^2)}{\vep_{k'}}[1+2n_b(\vep_{k'})],
\label{eq:r1}
\end{align}
where we used Wick's theorem and the symmetry of $\Gamma_4$ to simplify $\ac{\Pi_{i=1}^6\sigma_{\vk{i}}}=12 \ao{\si{1}\si{3}}\ao{\si{2}\si{4}}\ao{\si{5}\si{6}}$.
The remaining term of \eq{eq:tbc} to be evaluated is $-(1/4)\ac{( S_1)^2( S_3)^2}$. For this with expand the term $\ac{( S_1)^2( S_3)^2}=\ac{( S_1)^2( S_3^\sigma)^2}+\ac{( S_1)^2( S_3^\theta)^2}+2\ac{( S_1)^2 S_3^\theta S_3^\sigma}$, and evaluate each terms individually. They are given by
\begin{align}
-\dfrac{1}{4}\ac{( S_1)^2( S_3^\sigma)^2}
=&-\dfrac{1}{4}\sum_{\vk{1},..,\vk{8}}\Gi{1}\Gi{2}\Gs{3}{4}{5}\Gs{6}{7}{8}\ac{\Pi_{i=1}^8\sigma_{\vk{i}}}\delta_{\om_1,0}\delta_{\om_2,0}\delta_{\vk{3}+\vk{4}+\vk{5},0}\delta_{\vk{6}+\vk{7}+\vk{8},0}\notag\\
=&-\beta\dfrac{G^2\hbar^{12}}{32m^6}\int \dfrac{dk}{2\pi}\cos^2\left(\dfrac{k\ell}{2}\right)\dfrac{k^4}{\vep_k^4}\int_{|k'|<\Lambda} \dfrac{dk'}{2\pi}(k^2+k'^2-k k')^2 \dfrac{k'^2(k-k')^2}{\vep_{k'}\vep_{k-k'}}\notag\\
&\times\left\{\dfrac{\vep_{k-k'}[1+2n_b(\vep_{k'})]}{\vep_{k-k'}^2-\vep_{k'}^2}-  \dfrac{\vep_{k'}[1+2n_b(\vep_{k-k'})]}{\vep_{k-k'}^2-\vep_{k'}^2} \right\},
\label{eq:t1}
\end{align}
\begin{align}
-\dfrac{1}{4}\ac{( S_1)^2( S_3^\theta)^2}
=&-\dfrac{1}{4}\sum_{\vk{1},..,\vk{8}}\Gi{1}\Gi{2}\Gt{4}{5}\Gt{7}{8}\ac{\si{1}\si{2}\si{3}\te{4}\te{5}\si{6}\te{7}\te{8}}\delta_{\om_1,0}\delta_{\om_2,0}\notag\\	&\times\delta_{\vk{3}+\vk{4}+\vk{5},0}\delta_{\vk{6}+\vk{7}+\vk{8},0}\notag\\
=&-\beta\dfrac{G^2\hbar^4}{2m^2}\int \dfrac{dk}{2\pi}\cos^2\left(\dfrac{k\ell}{2}\right)\dfrac{k^4}{\vep_k^4}\int_{|k'|<\Lambda} \dfrac{dk'}{2\pi}\vep_{k'}\vep_{k-k'}\left\{\dfrac{\vep_{k-k'}[1+2n_b(\vep_{k'})]}{\vep_{k-k'}^2-\vep_{k'}^2}-  \dfrac{\vep_{k'}[1+2n_b(\vep_{k-k'})]}{\vep_{k-k'}^2-\vep_{k'}^2} \right\},\label{eq:t2}
\end{align}
\begin{align}
-\dfrac{1}{2}\ac{( S_1)^2 S_3^\theta S_3^\sigma}=&-\dfrac{1}{2}\sum_{\vk{1},..,\vk{8}}\Gi{1}\Gi{2}\Gt{4}{5}\Gs{6}{7}{8} \ac{\si{1}\si{2}\si{3}\te{4}\te{5}\si{6}\si{7}\si{8}}\delta_{\om_1,0}\delta_{\om_2,0}\notag\\
&\times\delta_{\vk{3}+\vk{4}+\vk{5},0}\delta_{\vk{6}+\vk{7}+\vk{8},0}
=\beta\dfrac{G^2\hbar^8}{4m^4}\int \dfrac{dk}{2\pi}\cos^2\left(\dfrac{k\ell}{2}\right)\dfrac{k^4}{\vep_k^4}\int_{|k'|<\Lambda} \dfrac{dk'}{2\pi}(k-k')k'(k^2+k'^2-k k')\notag\\
&\times\left\{\dfrac{\vep_{k-k'}[1+2n_b(\vep_{k-k'})]}{\vep_{k-k'}^2-\vep_{k'}^2}-  \dfrac{\vep_{k'}[1+2n_b(\vep_{k'})]}{\vep_{k-k'}^2-\vep_{k'}^2} \right\},\label{eq:t3}
\end{align}
where using Wick's theorem and the symmetries of $\Gamma_3^\sigma$ and $\Gamma_3^\theta$, we had the following simplifications: $\ac{\Pi_{i=1}^8\sigma_{\vk{i}}}=36 \ao{\si{1}\si{3}}\ao{\si{2}\si{6}}\ao{\si{4}\si{7}}\ao{\si{5}\si{8}}$ in \eq{eq:t1}, $\ac{\si{1}\si{2}\si{3}\te{4}\te{5}\si{6}\te{7}\te{8}}= 4\ao{\si{1}\si{3}}\ao{\si{2}\si{6}}\ao{\te{4}\te{7}}\ao{\te{5}\te{8}}$ in \eq{eq:t2}, and $\ac{\si{1}\si{2}\si{3}\te{4}\te{5}\si{6}\si{7}\si{8}}= 12\ao{\si{1}\si{3}}\ao{\si{2}\si{6}} \ao{\te{4}\si{7}}\ao{\te{5}\si{8}}$ in \eq{eq:t3}. Summing contributions (\ref{eq:r1}) to (\ref{eq:t3}), one obtains \eq{eq:tbc} from the main text, where
\begin{align}
\mathcal F^{(0)}(\ell)={}&\int \dfrac{dk}{2\pi} \cos^2\left(\dfrac{k \ell}{2} \right)\dfrac{k^4}{\vep_k^4}\int \dfrac{dk'}{2\pi} \biggl\{\dfrac{\hbar^{10}}{4m^5} \dfrac{k'^2(k^2+k'^2)}{\vep_{k'}}-\dfrac{\hbar^8 k^2}{4m^4} -\dfrac{\hbar^6}{2m^3}\dfrac{k'^2(k-k')^2}{\vep_{k'}\vep_{k-k'}}\dfrac{1}{\vep_{k'}+\vep_{k-k'}}\notag\\
&\times\left[\dfrac{\vep_{k'}\vep_{k-k'}}{k'(k-k')}-\dfrac{\hbar^4}{4m^2}(k^2+k'^2-kk') \right]^2\biggr\},\\
\mathcal F^{(T)}(\ell)={}&\int \dfrac{dk}{2\pi} \cos^2\left(\dfrac{k \ell}{2} \right)\dfrac{k^4}{\vep_k^4} \int \dfrac{dk'}{2\pi}\biggl\{\dfrac{\hbar^{10}}{2m^5} \dfrac{k'^2(k^2+k'^2)}{\vep_{k'}}n_b(\vep_{k'})+\dfrac{\hbar^{10}}{2m^5}k'(k-k')\dfrac{\vep_{k-k'}n_b(\vep_{k-k'})-\vep_{k'}n_b(\vep_{k'})}{\vep_{k-k'}^2-\vep_{k'}^2}\notag\\
&\times (k^2+k'^2-kk')-\dfrac{\hbar^6}{m^3}\left[\vep_{k'}\vep_{k-k'}+\dfrac{\hbar^8}{16m^4} \dfrac{k'^2(k-k')^2}{\vep_{k'}\vep_{k-k'}}(k^2+k'^2-kk')^2\right]\dfrac{\vep_{k-k'}n_b(\vep_{k'})-\vep_{k'}n_b(\vep_{k-k'})}{\vep_{k-k'}^2-\vep_{k'}^2}\biggr\}.
\end{align}
\end{widetext}	
By $n_b(\vep_k)=1/(e^{\beta \vep_k}-1)$ we denote the Bose-Einstein distribution function. After one integration one obtains the result  given in the main text.

\section{Alternative expression for the effective interaction at zero temperature}
\label{app:expre}

In our recent preprint \cite{reichert_casimir_2018}, we obtained the induced interaction between impurities at zero temperature using the approach based on the Gross-Pitaevskii equation where we accounted for the effect of quantum fluctuations. Within this formalism, the interaction can easily be split in a long-range and an exponentially decaying contribution as
\begin{align}
U^{(0)}(\ell)=U_\text{lr}(\ell)+U_\text{exp}(\ell). \label{eq:U0b}
\end{align}
The long-range part is
\begin{subequations}\label{C1}
\begin{align}
U_{\text{lr}}(\ell)=\dfrac{G^2 m}{2\pi\hbar^2}\int_{0}^{\infty} dx \frac{x^2\sin(2\ell x/\xi)}{\left(2+x^2\right)\sqrt{4+x^2}},
\label{eq:FLR}
\end{align}
while the exponentially decaying contribution is
\begin{align}
U_\text{exp}(\ell)=-\frac{G^2 m}{\hbar^2 \sqrt{\gamma}}e^{-2\ell /\xi}\left\{1+\frac{\sqrt{\gamma}}{2\pi}\left[1-\frac{2\ell}{\xi}+J_1\left(\frac{\ell}{\xi}\right)\right]\right\},\\
J_1(z)=\int_0^\infty dx\left[ \frac{2}{x\sqrt{4+x^2}}-\frac{8e^{-z(\sqrt{4+x^2}-2)}\cos(z x)}{x(2+x^2)(4+x^2)}\right].
\label{eq:uexp}
\end{align}
\end{subequations}
We notice that $J_1(z)$ is a monotonically increasing function with $J_1(0)=\ln 4$, while $J_1(z)=\ln z$ at large $z$. We were able to verify that the expression for the effective interaction  (\ref{eq:U0b}) is identical to the result (\ref{eq:U0}) of the main text.

The long-range part of the interaction (\ref{eq:FLR}) has  an asymptotic expansion
\begin{align}
U_{\text{lr}}(\ell)=&-\dfrac{ G^2m}{32\pi\hbar^2}\frac{\xi^3}{\ell^3}\sum_{n=0}^\infty \left(\frac{\xi}{2\ell}\right)^{2n} \biggl[\frac{\Gamma(2n+3)}{2^{n+1/2}}\notag\\
&- \frac{\Gamma(n+3/2)^2}{\pi} {_2F_1}\left(1,n+\frac{3}{2};
n+2;\frac{1}{2}\right)\biggr].
\end{align}
Here the hypergeometric function is defined as
\begin{align}\label{hyperg-def}
_2F_1(a_1,a_2;b_1;z)=\sum_{k=0}^\infty \frac{(a_1)_k (a_2)_k}{(b_1)_k}\frac{z^k}{k!},
\end{align}
where $(a)_k=\Gamma(a+k)/\Gamma(a)$. One can now easily obtain 
\begin{align}
U_\text{lr}(\ell)=&-\dfrac{ G^2m}{32\pi\hbar^2}\dfrac{\xi^3}{\ell^3}\left[1+\dfrac{15 \xi^2}{8\ell^2}+\dfrac{1935 \xi^4}{256 \ell^4} \right. \notag\\
&\left.+\dfrac{55755\xi^6}{1024\ell^6}+\dfrac{40639725\xi^8}{65536\ell^8}+\mathcal O(\xi^{10}/\ell^{10}) \right].\label{as}
\end{align}
We notice that the same asymptotic expansion (\ref{as}) is also obtained from Eq.~(\ref{eq:int1}). Let us mention that the two expressions for long-range interaction have identical asymptotic series, but they are not identical.

At small distances, $\ell\to 0$, naively one would use  $U_\text{lr}(0)=0$. However the limit does not commute with the integral and the correct result is $U_\text{lr}(\ell\to 0)=G^2m/4\hbar^2$, which leads to
\begin{align}
U^{(0)}(\ell\to 0)=-\frac{G^2 m}{\hbar^2\sqrt{\gamma}}\left[1+\frac{\sqrt{\gamma}}{\pi}\left(\ln 2+\frac{1}{2}-\frac{\pi}{4}\right)\right].
\end{align}
We obtained the same result using Eq.~(\ref{eq:U0}). 

\section{Bose gas in the absence of impurities}
\label{app:G0}

In the absence of impurity, i.e., for $G=0$, the Hamiltonian (\ref{eq:h0}) corresponds to the integrable Lieb-Liniger model \cite{lieb_exact_1963}. Using the formalism of the main text here we provide several thermodynamic quantities to one-loop order at small temperature and compare them with the known results in the literature. The Landau free energy  reads
\be
{F(\mu)}=-\dfrac{L\mu^2}{2g}-\dfrac{L\sqrt{m\mu^3}}{\pi \hbar}\left[\dfrac{2}{3}-f_0(\beta \mu)\right],\label{eq:FG0}
\ee
which enables us to obtain the density of the Bose gas
\be
\bar n(\mu)=-\dfrac{\partial F(\mu)}{L\partial \mu}=\dfrac{\mu}{g}+\dfrac{\sqrt{m\mu}}{\pi\hbar}\left[1+\dfrac{f_1(\beta \mu)}{2} \right]. \label{eq:n0G0}
\ee
We remind the reader that $f_0$ and $f_1$ are given by Eqs.~(\ref{eq:tf0}) and (\ref{eq:tf1}), respectively, and $\beta=1/T$.
Inverting the density (\ref{eq:n0G0}), one obtains the chemical potential
\begin{align}
\mu(\bar n)=\dfrac{\hbar^2\bar n^2\gamma}{m}\left\{1-\dfrac{\sqrt{\gamma}}{\pi}\left[1+\dfrac{1}{2}f_1\left(\dfrac{\hbar^2\bar n^2\gamma}{m T}\right) \right] \right\}	,	\label{eq:munG0}
\end{align}
where  $\gamma=g m/\hbar^2\bar n$.
From the chemical potential one can obtain the sound velocity
\be
v=\sqrt{\dfrac{\bar n}{m}\dfrac{\partial \mu|_{T=0}}{\partial \bar n}}=\dfrac{\hbar \bar n\sqrt{\gamma}}{m}\left(1-\dfrac{\sqrt{\gamma}}{4\pi} \right). \label{eq:vG0}
\ee
Performing the Legendre transformation (\ref{eq:en}) of \eq{eq:FG0} yields the Helmholtz free energy
\be
A(\bar n)=\dfrac{L \hbar^2\bar n^3\gamma}{m}\left\{\dfrac{1}{2}-\dfrac{\sqrt{\gamma}}{\pi}\left[\dfrac{2}{3}-f_0\left(\dfrac{\hbar^2\bar n^2\gamma}{m T}\right)\right]\right\}.\label{eq:EG0}
\ee
At zero temperature, $f_0=f_1=0$, and  Eqs.~(\ref{eq:munG0}) --(\ref{eq:EG0}) are in agreement with Refs.~\cite{lieb_exact_1963,popov_theory_1977}.
At small temperature $T\ll  \hbar^2\bar n^2\gamma/m$, one has
\be
f_{0,1}\left(\dfrac{\hbar^2\bar n^2\gamma}{m T}\right)= -\dfrac{\pi^2T^2m^2}{6\hbar^4\bar n^4\gamma^2}.
\ee
In this limit, the chemical potential and Helmholtz free energy read
\begin{align}
\mu(\bar n)={}&\dfrac{\hbar^2\bar n^2\gamma}{m}\left(1-\dfrac{\sqrt{\gamma}}{\pi}\right)+\dfrac{\pi m T^2}{12\hbar^2\bar n^2 \sqrt{\gamma}},\\
A(\bar n)={}&\dfrac{L \hbar^2\bar n^3\gamma}{2m}\left(1-\dfrac{4\sqrt{\gamma}}{3\pi}\right)-\dfrac{\pi L m  T^2}{6\hbar^2 \bar n \sqrt{\gamma}}.
\end{align}
The last two results are in agreement with Ref.~\cite{de_rosi_thermodynamic_2017}.

%\bibliography{bibQFT}

\begin{thebibliography}{36}%
	\makeatletter
	\providecommand \@ifxundefined [1]{%
		\@ifx{#1\undefined}
	}%
	\providecommand \@ifnum [1]{%
		\ifnum #1\expandafter \@firstoftwo
		\else \expandafter \@secondoftwo
		\fi
	}%
	\providecommand \@ifx [1]{%
		\ifx #1\expandafter \@firstoftwo
		\else \expandafter \@secondoftwo
		\fi
	}%
	\providecommand \natexlab [1]{#1}%
	\providecommand \enquote  [1]{``#1''}%
	\providecommand \bibnamefont  [1]{#1}%
	\providecommand \bibfnamefont [1]{#1}%
	\providecommand \citenamefont [1]{#1}%
	\providecommand \href@noop [0]{\@secondoftwo}%
	\providecommand \href [0]{\begingroup \@sanitize@url \@href}%
	\providecommand \@href[1]{\@@startlink{#1}\@@href}%
	\providecommand \@@href[1]{\endgroup#1\@@endlink}%
	\providecommand \@sanitize@url [0]{\catcode `\\12\catcode `\$12\catcode
		`\&12\catcode `\#12\catcode `\^12\catcode `\_12\catcode `\%12\relax}%
	\providecommand \@@startlink[1]{}%
	\providecommand \@@endlink[0]{}%
	\providecommand \url  [0]{\begingroup\@sanitize@url \@url }%
	\providecommand \@url [1]{\endgroup\@href {#1}{\urlprefix }}%
	\providecommand \urlprefix  [0]{URL }%
	\providecommand \Eprint [0]{\href }%
	\providecommand \doibase [0]{http://dx.doi.org/}%
	\providecommand \selectlanguage [0]{\@gobble}%
	\providecommand \bibinfo  [0]{\@secondoftwo}%
	\providecommand \bibfield  [0]{\@secondoftwo}%
	\providecommand \translation [1]{[#1]}%
	\providecommand \BibitemOpen [0]{}%
	\providecommand \bibitemStop [0]{}%
	\providecommand \bibitemNoStop [0]{.\EOS\space}%
	\providecommand \EOS [0]{\spacefactor3000\relax}%
	\providecommand \BibitemShut  [1]{\csname bibitem#1\endcsname}%
	\let\auto@bib@innerbib\@empty
	%</preamble>
	\bibitem [{\citenamefont {Casimir}(1948)}]{casimir1948}%
	\BibitemOpen
	\bibfield  {author} {\bibinfo {author} {\bibfnamefont {H.~B.~G.}\
			\bibnamefont {Casimir}},\ }\href@noop {} {\bibfield  {journal} {\bibinfo
			{journal} {Proc. K. Ned. Akad. Wet.}\ }\textbf {\bibinfo {volume} {51}},\
		\bibinfo {pages} {793} (\bibinfo {year} {1948})}\BibitemShut {NoStop}%
	\bibitem [{\citenamefont {Bordag}\ \emph {et~al.}(2009)\citenamefont {Bordag},
		\citenamefont {Klimchitskaya}, \citenamefont {Mohideen},\ and\ \citenamefont
		{Mostepanenko}}]{bordag_advances_2009}%
	\BibitemOpen
	\bibfield  {author} {\bibinfo {author} {\bibfnamefont {M.}~\bibnamefont
			{Bordag}}, \bibinfo {author} {\bibfnamefont {G.~L.}\ \bibnamefont
			{Klimchitskaya}}, \bibinfo {author} {\bibfnamefont {U.}~\bibnamefont
			{Mohideen}}, \ and\ \bibinfo {author} {\bibfnamefont {V.~M.}\ \bibnamefont
			{Mostepanenko}},\ }\href@noop {} {\emph {\bibinfo {title} {Advances in the
				{Casimir} {Effect}}}},\ International {Series} of {Monographs} on {Physics}\
	(\bibinfo  {publisher} {Oxford University Press},\ \bibinfo {address}
	{Oxford, New York},\ \bibinfo {year} {2009})\BibitemShut {NoStop}%
	\bibitem [{\citenamefont {Lamoreaux}(1997)}]{lamoreaux_demonstration_1997}%
	\BibitemOpen
	\bibfield  {author} {\bibinfo {author} {\bibfnamefont {S.~K.}\ \bibnamefont
			{Lamoreaux}},\ }\href {\doibase 10.1103/PhysRevLett.78.5} {\bibfield
		{journal} {\bibinfo  {journal} {Physical Review Letters}\ }\textbf {\bibinfo
			{volume} {78}},\ \bibinfo {pages} {5} (\bibinfo {year} {1997})}\BibitemShut
	{NoStop}%
	\bibitem [{\citenamefont {Mohideen}\ and\ \citenamefont
		{Roy}(1998)}]{mohideen_precision_1998}%
	\BibitemOpen
	\bibfield  {author} {\bibinfo {author} {\bibfnamefont {U.}~\bibnamefont
			{Mohideen}}\ and\ \bibinfo {author} {\bibfnamefont {A.}~\bibnamefont {Roy}},\
	}\href {\doibase 10.1103/PhysRevLett.81.4549} {\bibfield  {journal} {\bibinfo
			{journal} {Physical Review Letters}\ }\textbf {\bibinfo {volume} {81}},\
		\bibinfo {pages} {4549} (\bibinfo {year} {1998})}\BibitemShut {NoStop}%
	\bibitem [{\citenamefont {Klimchitskaya}\ \emph {et~al.}(2009)\citenamefont
		{Klimchitskaya}, \citenamefont {Mohideen},\ and\ \citenamefont
		{Mostepanenko}}]{klimchitskaya_casimir_2009}%
	\BibitemOpen
	\bibfield  {author} {\bibinfo {author} {\bibfnamefont {G.~L.}\ \bibnamefont
			{Klimchitskaya}}, \bibinfo {author} {\bibfnamefont {U.}~\bibnamefont
			{Mohideen}}, \ and\ \bibinfo {author} {\bibfnamefont {V.~M.}\ \bibnamefont
			{Mostepanenko}},\ }\href {\doibase 10.1103/RevModPhys.81.1827} {\bibfield
		{journal} {\bibinfo  {journal} {Reviews of Modern Physics}\ }\textbf
		{\bibinfo {volume} {81}},\ \bibinfo {pages} {1827} (\bibinfo {year}
		{2009})}\BibitemShut {NoStop}%
	\bibitem [{\citenamefont {Kardar}\ and\ \citenamefont
		{Golestanian}(1999)}]{kardar_friction_1999}%
	\BibitemOpen
	\bibfield  {author} {\bibinfo {author} {\bibfnamefont {M.}~\bibnamefont
			{Kardar}}\ and\ \bibinfo {author} {\bibfnamefont {R.}~\bibnamefont
			{Golestanian}},\ }\href {\doibase 10.1103/RevModPhys.71.1233} {\bibfield
		{journal} {\bibinfo  {journal} {Reviews of Modern Physics}\ }\textbf
		{\bibinfo {volume} {71}},\ \bibinfo {pages} {1233} (\bibinfo {year}
		{1999})}\BibitemShut {NoStop}%
	\bibitem [{\citenamefont {Bloch}\ \emph {et~al.}(2008)\citenamefont {Bloch},
		\citenamefont {Dalibard},\ and\ \citenamefont
		{Zwerger}}]{bloch_many-body_2008}%
	\BibitemOpen
	\bibfield  {author} {\bibinfo {author} {\bibfnamefont {I.}~\bibnamefont
			{Bloch}}, \bibinfo {author} {\bibfnamefont {J.}~\bibnamefont {Dalibard}}, \
		and\ \bibinfo {author} {\bibfnamefont {W.}~\bibnamefont {Zwerger}},\ }\href
	{\doibase 10.1103/RevModPhys.80.885} {\bibfield  {journal} {\bibinfo
			{journal} {Reviews of Modern Physics}\ }\textbf {\bibinfo {volume} {80}},\
		\bibinfo {pages} {885} (\bibinfo {year} {2008})}\BibitemShut {NoStop}%
	\bibitem [{\citenamefont {Bardeen}\ \emph {et~al.}(1967)\citenamefont
		{Bardeen}, \citenamefont {Baym},\ and\ \citenamefont
		{Pines}}]{bardeen_effective_1967}%
	\BibitemOpen
	\bibfield  {author} {\bibinfo {author} {\bibfnamefont {J.}~\bibnamefont
			{Bardeen}}, \bibinfo {author} {\bibfnamefont {G.}~\bibnamefont {Baym}}, \
		and\ \bibinfo {author} {\bibfnamefont {D.}~\bibnamefont {Pines}},\ }\href
	{\doibase 10.1103/PhysRev.156.207} {\bibfield  {journal} {\bibinfo  {journal}
			{Physical Review}\ }\textbf {\bibinfo {volume} {156}},\ \bibinfo {pages}
		{207} (\bibinfo {year} {1967})}\BibitemShut {NoStop}%
	\bibitem [{\citenamefont {Bijlsma}\ \emph {et~al.}(2000)\citenamefont
		{Bijlsma}, \citenamefont {Heringa},\ and\ \citenamefont
		{Stoof}}]{bijlsma_phonon_2000}%
	\BibitemOpen
	\bibfield  {author} {\bibinfo {author} {\bibfnamefont {M.~J.}\ \bibnamefont
			{Bijlsma}}, \bibinfo {author} {\bibfnamefont {B.~A.}\ \bibnamefont
			{Heringa}}, \ and\ \bibinfo {author} {\bibfnamefont {H.~T.~C.}\ \bibnamefont
			{Stoof}},\ }\href {\doibase 10.1103/PhysRevA.61.053601} {\bibfield  {journal}
		{\bibinfo  {journal} {Physical Review A}\ }\textbf {\bibinfo {volume} {61}},\
		\bibinfo {pages} {053601} (\bibinfo {year} {2000})}\BibitemShut {NoStop}%
	\bibitem [{\citenamefont {Roberts}\ and\ \citenamefont
		{Pomeau}(2005)}]{roberts_casimir-like_2005}%
	\BibitemOpen
	\bibfield  {author} {\bibinfo {author} {\bibfnamefont {D.~C.}\ \bibnamefont
			{Roberts}}\ and\ \bibinfo {author} {\bibfnamefont {Y.}~\bibnamefont
			{Pomeau}},\ }\href {\doibase 10.1103/PhysRevLett.95.145303} {\bibfield
		{journal} {\bibinfo  {journal} {Physical Review Letters}\ }\textbf {\bibinfo
			{volume} {95}},\ \bibinfo {pages} {145303} (\bibinfo {year}
		{2005})}\BibitemShut {NoStop}%
	\bibitem [{\citenamefont {Klein}\ and\ \citenamefont
		{Fleischhauer}(2005)}]{klein_interaction_2005}%
	\BibitemOpen
	\bibfield  {author} {\bibinfo {author} {\bibfnamefont {A.}~\bibnamefont
			{Klein}}\ and\ \bibinfo {author} {\bibfnamefont {M.}~\bibnamefont
			{Fleischhauer}},\ }\href {\doibase 10.1103/PhysRevA.71.033605} {\bibfield
		{journal} {\bibinfo  {journal} {Physical Review A}\ }\textbf {\bibinfo
			{volume} {71}},\ \bibinfo {pages} {033605} (\bibinfo {year}
		{2005})}\BibitemShut {NoStop}%
	\bibitem [{\citenamefont {Recati}\ \emph {et~al.}(2005)\citenamefont {Recati},
		\citenamefont {Fuchs}, \citenamefont {Peça},\ and\ \citenamefont
		{Zwerger}}]{recati_casimir_2005}%
	\BibitemOpen
	\bibfield  {author} {\bibinfo {author} {\bibfnamefont {A.}~\bibnamefont
			{Recati}}, \bibinfo {author} {\bibfnamefont {J.~N.}\ \bibnamefont {Fuchs}},
		\bibinfo {author} {\bibfnamefont {C.~S.}\ \bibnamefont {Peça}}, \ and\
		\bibinfo {author} {\bibfnamefont {W.}~\bibnamefont {Zwerger}},\ }\href
	{\doibase 10.1103/PhysRevA.72.023616} {\bibfield  {journal} {\bibinfo
			{journal} {Physical Review A}\ }\textbf {\bibinfo {volume} {72}},\ \bibinfo
		{pages} {023616} (\bibinfo {year} {2005})}\BibitemShut {NoStop}%
	\bibitem [{\citenamefont {Fuchs}\ \emph {et~al.}(2007)\citenamefont {Fuchs},
		\citenamefont {Recati},\ and\ \citenamefont
		{Zwerger}}]{fuchs_oscillating_2007}%
	\BibitemOpen
	\bibfield  {author} {\bibinfo {author} {\bibfnamefont {J.~N.}\ \bibnamefont
			{Fuchs}}, \bibinfo {author} {\bibfnamefont {A.}~\bibnamefont {Recati}}, \
		and\ \bibinfo {author} {\bibfnamefont {W.}~\bibnamefont {Zwerger}},\ }\href
	{\doibase 10.1103/PhysRevA.75.043615} {\bibfield  {journal} {\bibinfo
			{journal} {Physical Review A}\ }\textbf {\bibinfo {volume} {75}},\ \bibinfo
		{pages} {043615} (\bibinfo {year} {2007})}\BibitemShut {NoStop}%
	\bibitem [{\citenamefont {Wächter}\ \emph {et~al.}(2007)\citenamefont
		{Wächter}, \citenamefont {Meden},\ and\ \citenamefont
		{Schönhammer}}]{wachter_indirect_2007}%
	\BibitemOpen
	\bibfield  {author} {\bibinfo {author} {\bibfnamefont {P.}~\bibnamefont
			{Wächter}}, \bibinfo {author} {\bibfnamefont {V.}~\bibnamefont {Meden}}, \
		and\ \bibinfo {author} {\bibfnamefont {K.}~\bibnamefont {Schönhammer}},\
	}\href {\doibase 10.1103/PhysRevB.76.045123} {\bibfield  {journal} {\bibinfo
			{journal} {Physical Review B}\ }\textbf {\bibinfo {volume} {76}},\ \bibinfo
		{pages} {045123} (\bibinfo {year} {2007})}\BibitemShut {NoStop}%
	\bibitem [{\citenamefont {Schecter}\ and\ \citenamefont
		{Kamenev}(2014)}]{schecter_phonon-mediated_2014}%
	\BibitemOpen
	\bibfield  {author} {\bibinfo {author} {\bibfnamefont {M.}~\bibnamefont
			{Schecter}}\ and\ \bibinfo {author} {\bibfnamefont {A.}~\bibnamefont
			{Kamenev}},\ }\href {\doibase 10.1103/PhysRevLett.112.155301} {\bibfield
		{journal} {\bibinfo  {journal} {Physical Review Letters}\ }\textbf {\bibinfo
			{volume} {112}},\ \bibinfo {pages} {155301} (\bibinfo {year}
		{2014})}\BibitemShut {NoStop}%
	\bibitem [{\citenamefont {Marino}\ \emph {et~al.}(2017)\citenamefont {Marino},
		\citenamefont {Recati},\ and\ \citenamefont
		{Carusotto}}]{marino_casimir_2017}%
	\BibitemOpen
	\bibfield  {author} {\bibinfo {author} {\bibfnamefont {J.}~\bibnamefont
			{Marino}}, \bibinfo {author} {\bibfnamefont {A.}~\bibnamefont {Recati}}, \
		and\ \bibinfo {author} {\bibfnamefont {I.}~\bibnamefont {Carusotto}},\ }\href
	{\doibase 10.1103/PhysRevLett.118.045301} {\bibfield  {journal} {\bibinfo
			{journal} {Physical Review Letters}\ }\textbf {\bibinfo {volume} {118}},\
		\bibinfo {pages} {045301} (\bibinfo {year} {2017})}\BibitemShut {NoStop}%
	\bibitem [{\citenamefont {Dehkharghani}\ \emph {et~al.}(2018)\citenamefont
		{Dehkharghani}, \citenamefont {Volosniev},\ and\ \citenamefont
		{Zinner}}]{dehkharghani_coalescence_2018}%
	\BibitemOpen
	\bibfield  {author} {\bibinfo {author} {\bibfnamefont {A.}~\bibnamefont
			{Dehkharghani}}, \bibinfo {author} {\bibfnamefont {A.}~\bibnamefont
			{Volosniev}}, \ and\ \bibinfo {author} {\bibfnamefont {N.}~\bibnamefont
			{Zinner}},\ }\href {\doibase 10.1103/PhysRevLett.121.080405} {\bibfield
		{journal} {\bibinfo  {journal} {Physical Review Letters}\ }\textbf {\bibinfo
			{volume} {121}},\ \bibinfo {pages} {080405} (\bibinfo {year}
		{2018})}\BibitemShut {NoStop}%
	\bibitem [{\citenamefont {Yu}\ \emph {et~al.}(2009)\citenamefont {Yu},
		\citenamefont {Qi}, \citenamefont {Li},\ and\ \citenamefont
		{Liu}}]{yu_casimir_2009}%
	\BibitemOpen
	\bibfield  {author} {\bibinfo {author} {\bibfnamefont {X.-L.}\ \bibnamefont
			{Yu}}, \bibinfo {author} {\bibfnamefont {R.}~\bibnamefont {Qi}}, \bibinfo
		{author} {\bibfnamefont {Z.~B.}\ \bibnamefont {Li}}, \ and\ \bibinfo {author}
		{\bibfnamefont {W.~M.}\ \bibnamefont {Liu}},\ }\href {\doibase
		10.1209/0295-5075/85/10005} {\bibfield  {journal} {\bibinfo  {journal}
			{Europhysics Letters}\ }\textbf {\bibinfo {volume} {85}},\ \bibinfo {pages}
		{10005} (\bibinfo {year} {2009})}\BibitemShut {NoStop}%
	\bibitem [{\citenamefont {Reichert}\ \emph {et~al.}(2019)\citenamefont
		{Reichert}, \citenamefont {Ristivojevic},\ and\ \citenamefont
		{Petkovic}}]{reichert_casimir_2018}%
	\BibitemOpen
	\bibfield  {author} {\bibinfo {author} {\bibfnamefont {B.}~\bibnamefont
			{Reichert}}, \bibinfo {author} {\bibfnamefont {Z.}~\bibnamefont
			{Ristivojevic}}, \ and\ \bibinfo {author} {\bibfnamefont {A.}~\bibnamefont
			{Petkovic}},\ }\href {https://doi.org/10.1088/1367-2630/ab1b8e} {\bibfield
		{journal} {\bibinfo  {journal} {New Journal of Physics}\ } (\bibinfo {year}
		{2019})}\BibitemShut {NoStop}%
	\bibitem [{\citenamefont {Pavlov}\ \emph {et~al.}(2018)\citenamefont {Pavlov},
		\citenamefont {van~den Brink},\ and\ \citenamefont
		{Efremov}}]{pavlov_phonon-mediated_2018}%
	\BibitemOpen
	\bibfield  {author} {\bibinfo {author} {\bibfnamefont {A.~I.}\ \bibnamefont
			{Pavlov}}, \bibinfo {author} {\bibfnamefont {J.}~\bibnamefont {van~den
				Brink}}, \ and\ \bibinfo {author} {\bibfnamefont {D.~V.}\ \bibnamefont
			{Efremov}},\ }\href {\doibase 10.1103/PhysRevB.98.161410} {\bibfield
		{journal} {\bibinfo  {journal} {Physical Review B}\ }\textbf {\bibinfo
			{volume} {98}},\ \bibinfo {pages} {161410} (\bibinfo {year}
		{2018})}\BibitemShut {NoStop}%
	\bibitem [{\citenamefont {Lieb}\ and\ \citenamefont
		{Liniger}(1963)}]{lieb_exact_1963}%
	\BibitemOpen
	\bibfield  {author} {\bibinfo {author} {\bibfnamefont {E.~H.}\ \bibnamefont
			{Lieb}}\ and\ \bibinfo {author} {\bibfnamefont {W.}~\bibnamefont {Liniger}},\
	}\href {\doibase 10.1103/PhysRev.130.1605} {\bibfield  {journal} {\bibinfo
			{journal} {Physical Review}\ }\textbf {\bibinfo {volume} {130}},\ \bibinfo
		{pages} {1605} (\bibinfo {year} {1963})}\BibitemShut {NoStop}%
	\bibitem [{\citenamefont {Braaten}\ and\ \citenamefont
		{Nieto}(1999)}]{braaten_quantum_1999}%
	\BibitemOpen
	\bibfield  {author} {\bibinfo {author} {\bibfnamefont {E.}~\bibnamefont
			{Braaten}}\ and\ \bibinfo {author} {\bibfnamefont {A.}~\bibnamefont
			{Nieto}},\ }\href {\doibase 10.1007/s100510050925} {\bibfield  {journal}
		{\bibinfo  {journal} {The European Physical Journal B - Condensed Matter and
				Complex Systems}\ }\textbf {\bibinfo {volume} {11}},\ \bibinfo {pages} {143}
		(\bibinfo {year} {1999})}\BibitemShut {NoStop}%
	\bibitem [{\citenamefont {Andersen}(2004)}]{andersen_theory_2004}%
	\BibitemOpen
	\bibfield  {author} {\bibinfo {author} {\bibfnamefont {J.~O.}\ \bibnamefont
			{Andersen}},\ }\href {\doibase 10.1103/RevModPhys.76.599} {\bibfield
		{journal} {\bibinfo  {journal} {Reviews of Modern Physics}\ }\textbf
		{\bibinfo {volume} {76}},\ \bibinfo {pages} {599} (\bibinfo {year}
		{2004})}\BibitemShut {NoStop}%
	\bibitem [{\citenamefont {Gangardt}\ and\ \citenamefont
		{Kamenev}(2009)}]{gangardt_bloch_2009}%
	\BibitemOpen
	\bibfield  {author} {\bibinfo {author} {\bibfnamefont {D.~M.}\ \bibnamefont
			{Gangardt}}\ and\ \bibinfo {author} {\bibfnamefont {A.}~\bibnamefont
			{Kamenev}},\ }\href {\doibase 10.1103/PhysRevLett.102.070402} {\bibfield
		{journal} {\bibinfo  {journal} {Physical Review Letters}\ }\textbf {\bibinfo
			{volume} {102}},\ \bibinfo {pages} {070402} (\bibinfo {year}
		{2009})}\BibitemShut {NoStop}%
	\bibitem [{\citenamefont {Ristivojevic}\ and\ \citenamefont
		{Matveev}(2016)}]{ristivojevic_decay_2016}%
	\BibitemOpen
	\bibfield  {author} {\bibinfo {author} {\bibfnamefont {Z.}~\bibnamefont
			{Ristivojevic}}\ and\ \bibinfo {author} {\bibfnamefont {K.~A.}\ \bibnamefont
			{Matveev}},\ }\href {\doibase 10.1103/PhysRevB.94.024506} {\bibfield
		{journal} {\bibinfo  {journal} {Physical Review B}\ }\textbf {\bibinfo
			{volume} {94}},\ \bibinfo {pages} {024506} (\bibinfo {year}
		{2016})}\BibitemShut {NoStop}%
	\bibitem [{\citenamefont {Ma}(1993)}]{ma}%
	\BibitemOpen
	\bibfield  {author} {\bibinfo {author} {\bibfnamefont {S.-K.}\ \bibnamefont
			{Ma}},\ }\href@noop {} {\emph {\bibinfo {title} {Statistical Mechanics}}}\
	(\bibinfo  {publisher} {World Scientific},\ \bibinfo {year}
	{1993})\BibitemShut {NoStop}%
	\bibitem [{\citenamefont {Pustilnik}\ and\ \citenamefont
		{Matveev}(2014)}]{pustilnik_low-energy_2014}%
	\BibitemOpen
	\bibfield  {author} {\bibinfo {author} {\bibfnamefont {M.}~\bibnamefont
			{Pustilnik}}\ and\ \bibinfo {author} {\bibfnamefont {K.~A.}\ \bibnamefont
			{Matveev}},\ }\href {\doibase 10.1103/PhysRevB.89.100504} {\bibfield
		{journal} {\bibinfo  {journal} {Physical Review B}\ }\textbf {\bibinfo
			{volume} {89}},\ \bibinfo {pages} {100504} (\bibinfo {year}
		{2014})}\BibitemShut {NoStop}%
	\bibitem [{\citenamefont {Imambekov}\ \emph {et~al.}(2012)\citenamefont
		{Imambekov}, \citenamefont {Schmidt},\ and\ \citenamefont
		{Glazman}}]{imambekov_one-dimensional_2012}%
	\BibitemOpen
	\bibfield  {author} {\bibinfo {author} {\bibfnamefont {A.}~\bibnamefont
			{Imambekov}}, \bibinfo {author} {\bibfnamefont {T.~L.}\ \bibnamefont
			{Schmidt}}, \ and\ \bibinfo {author} {\bibfnamefont {L.~I.}\ \bibnamefont
			{Glazman}},\ }\href {\doibase 10.1103/RevModPhys.84.1253} {\bibfield
		{journal} {\bibinfo  {journal} {Reviews of Modern Physics}\ }\textbf
		{\bibinfo {volume} {84}},\ \bibinfo {pages} {1253} (\bibinfo {year}
		{2012})}\BibitemShut {NoStop}%
	\bibitem [{\citenamefont {Kheruntsyan}\ \emph {et~al.}(2003)\citenamefont
		{Kheruntsyan}, \citenamefont {Gangardt}, \citenamefont {Drummond},\ and\
		\citenamefont {Shlyapnikov}}]{kheruntsyan_pair_2003}%
	\BibitemOpen
	\bibfield  {author} {\bibinfo {author} {\bibfnamefont {K.~V.}\ \bibnamefont
			{Kheruntsyan}}, \bibinfo {author} {\bibfnamefont {D.~M.}\ \bibnamefont
			{Gangardt}}, \bibinfo {author} {\bibfnamefont {P.~D.}\ \bibnamefont
			{Drummond}}, \ and\ \bibinfo {author} {\bibfnamefont {G.~V.}\ \bibnamefont
			{Shlyapnikov}},\ }\href {\doibase 10.1103/PhysRevLett.91.040403} {\bibfield
		{journal} {\bibinfo  {journal} {Physical Review Letters}\ }\textbf {\bibinfo
			{volume} {91}},\ \bibinfo {pages} {040403} (\bibinfo {year}
		{2003})}\BibitemShut {NoStop}%
	\bibitem [{\citenamefont {Casimir}\ and\ \citenamefont
		{Polder}(1948)}]{casimir_influence_1948}%
	\BibitemOpen
	\bibfield  {author} {\bibinfo {author} {\bibfnamefont {H.~B.~G.}\
			\bibnamefont {Casimir}}\ and\ \bibinfo {author} {\bibfnamefont
			{D.}~\bibnamefont {Polder}},\ }\href {\doibase 10.1103/PhysRev.73.360}
	{\bibfield  {journal} {\bibinfo  {journal} {Physical Review}\ }\textbf
		{\bibinfo {volume} {73}},\ \bibinfo {pages} {360} (\bibinfo {year}
		{1948})}\BibitemShut {NoStop}%
	\bibitem [{\citenamefont {Hofferberth}\ \emph {et~al.}(2008)\citenamefont
		{Hofferberth}, \citenamefont {Lesanovsky}, \citenamefont {Schumm},
		\citenamefont {Imambekov}, \citenamefont {Gritsev}, \citenamefont {Demler},\
		and\ \citenamefont {Schmiedmayer}}]{hofferberth_probing_2008}%
	\BibitemOpen
	\bibfield  {author} {\bibinfo {author} {\bibfnamefont {S.}~\bibnamefont
			{Hofferberth}}, \bibinfo {author} {\bibfnamefont {I.}~\bibnamefont
			{Lesanovsky}}, \bibinfo {author} {\bibfnamefont {T.}~\bibnamefont {Schumm}},
		\bibinfo {author} {\bibfnamefont {A.}~\bibnamefont {Imambekov}}, \bibinfo
		{author} {\bibfnamefont {V.}~\bibnamefont {Gritsev}}, \bibinfo {author}
		{\bibfnamefont {E.}~\bibnamefont {Demler}}, \ and\ \bibinfo {author}
		{\bibfnamefont {J.}~\bibnamefont {Schmiedmayer}},\ }\href {\doibase
		10.1038/nphys941} {\bibfield  {journal} {\bibinfo  {journal} {Nature
				Physics}\ }\textbf {\bibinfo {volume} {4}},\ \bibinfo {pages} {489} (\bibinfo
		{year} {2008})}\BibitemShut {NoStop}%
	\bibitem [{\citenamefont {Roberts}\ and\ \citenamefont
		{Rica}(2009)}]{roberts_impurity_2009}%
	\BibitemOpen
	\bibfield  {author} {\bibinfo {author} {\bibfnamefont {D.~C.}\ \bibnamefont
			{Roberts}}\ and\ \bibinfo {author} {\bibfnamefont {S.}~\bibnamefont {Rica}},\
	}\href {\doibase 10.1103/PhysRevLett.102.025301} {\bibfield  {journal}
		{\bibinfo  {journal} {Physical Review Letters}\ }\textbf {\bibinfo {volume}
			{102}},\ \bibinfo {pages} {025301} (\bibinfo {year} {2009})}\BibitemShut
	{NoStop}%
	\bibitem [{\citenamefont {Casteels}\ \emph {et~al.}(2013)\citenamefont
		{Casteels}, \citenamefont {Tempere},\ and\ \citenamefont
		{Devreese}}]{casteels_bipolarons_2013}%
	\BibitemOpen
	\bibfield  {author} {\bibinfo {author} {\bibfnamefont {W.}~\bibnamefont
			{Casteels}}, \bibinfo {author} {\bibfnamefont {J.}~\bibnamefont {Tempere}}, \
		and\ \bibinfo {author} {\bibfnamefont {J.~T.}\ \bibnamefont {Devreese}},\
	}\href {\doibase 10.1103/PhysRevA.88.013613} {\bibfield  {journal} {\bibinfo
			{journal} {Physical Review A}\ }\textbf {\bibinfo {volume} {88}},\ \bibinfo
		{pages} {013613} (\bibinfo {year} {2013})}\BibitemShut {NoStop}%
	\bibitem [{\citenamefont {Camacho-Guardian}\ \emph {et~al.}(2018)\citenamefont
		{Camacho-Guardian}, \citenamefont {Peña~Ardila}, \citenamefont {Pohl},\ and\
		\citenamefont {Bruun}}]{camacho-guardian_bipolarons_2018}%
	\BibitemOpen
	\bibfield  {author} {\bibinfo {author} {\bibfnamefont {A.}~\bibnamefont
			{Camacho-Guardian}}, \bibinfo {author} {\bibfnamefont {L.}~\bibnamefont
			{Peña~Ardila}}, \bibinfo {author} {\bibfnamefont {T.}~\bibnamefont {Pohl}},
		\ and\ \bibinfo {author} {\bibfnamefont {G.}~\bibnamefont {Bruun}},\ }\href
	{\doibase 10.1103/PhysRevLett.121.013401} {\bibfield  {journal} {\bibinfo
			{journal} {Physical Review Letters}\ }\textbf {\bibinfo {volume} {121}},\
		\bibinfo {pages} {013401} (\bibinfo {year} {2018})}\BibitemShut {NoStop}%
	\bibitem [{\citenamefont {Popov}(1977)}]{popov_theory_1977}%
	\BibitemOpen
	\bibfield  {author} {\bibinfo {author} {\bibfnamefont {V.~N.}\ \bibnamefont
			{Popov}},\ }\href {\doibase 10.1007/BF01036714} {\bibfield  {journal}
		{\bibinfo  {journal} {Theoretical and Mathematical Physics}\ }\textbf
		{\bibinfo {volume} {30}},\ \bibinfo {pages} {222} (\bibinfo {year}
		{1977})}\BibitemShut {NoStop}%
	\bibitem [{\citenamefont {De~Rosi}\ \emph {et~al.}(2017)\citenamefont
		{De~Rosi}, \citenamefont {Astrakharchik},\ and\ \citenamefont
		{Stringari}}]{de_rosi_thermodynamic_2017}%
	\BibitemOpen
	\bibfield  {author} {\bibinfo {author} {\bibfnamefont {G.}~\bibnamefont
			{De~Rosi}}, \bibinfo {author} {\bibfnamefont {G.~E.}\ \bibnamefont
			{Astrakharchik}}, \ and\ \bibinfo {author} {\bibfnamefont {S.}~\bibnamefont
			{Stringari}},\ }\href {\doibase 10.1103/PhysRevA.96.013613} {\bibfield
		{journal} {\bibinfo  {journal} {Physical Review A}\ }\textbf {\bibinfo
			{volume} {96}},\ \bibinfo {pages} {013613} (\bibinfo {year}
		{2017})}\BibitemShut {NoStop}%
\end{thebibliography}


\end{document}